\documentclass[sigconf]{acmart}
\AtBeginDocument{%
  \providecommand\BibTeX{{%
    \normalfont B\kern-0.5em{\scshape i\kern-0.25em b}\kern-0.8em\TeX}}}

\renewcommand\footnotetextcopyrightpermission[1]{} % removes footnote with conference information in first column
\setcopyright{none}

\usepackage{xcolor}

\usepackage{algorithm}
\usepackage{algorithmic}
\usepackage{amsmath}

\usepackage{booktabs}
\usepackage{multirow} 
\usepackage{subcaption}
\usepackage{makecell}
\usepackage{tabularx}
\usepackage{enumitem}
\usepackage{listings}

\begin{document}

\title{Zero-Shot Character Identification and Speaker Prediction in Comics via Iterative Multimodal Fusion}


\author{Yingxuan Li}
\affiliation{
  \institution{The University of Tokyo}
  \city{Tokyo}
  \country{Japan}}
\email{li@hal.t.u-tokyo.ac.jp}
\thanks{*This work was conducted when the first author was a research intern at Mantra Inc. For more details, please visit our project page: \url{https://liyingxuan1012.github.io/zeroshot-speaker-prediction}.}

\author{Ryota Hinami}
\affiliation{
  \institution{Mantra Inc.}
  \city{Tokyo}
  \country{Japan}}
\email{hinami@mantra.co.jp}

\author{Kiyoharu Aizawa}
\affiliation{
  \institution{The University of Tokyo}
  \city{Tokyo}
  \country{Japan}}
\email{aizawa@hal.t.u-tokyo.ac.jp}

\author{Yusuke Matsui}
\affiliation{
  \institution{The University of Tokyo}
  \city{Tokyo}
  \country{Japan}}
\email{matsui@hal.t.u-tokyo.ac.jp}

% \author{Yingxuan Li$^1$,\hspace{1em} Ryota Hinami$^2$,\hspace{1em} Kiyoharu Aizawa$^1$,\hspace{1em} Yusuke Matsui$^1$}
% \thanks{*This work was conducted when the first author was a research intern at Mantra Inc. For more details, please visit our project page: \url{https://liyingxuan1012.github.io/ zeroshot- speaker- prediction.}}
% \affiliation{%
% \institution{$^1$The University of Tokyo \hspace{0.3em} $^2$Mantra Inc.}
% \city{}
% \country{}}
% \affiliation{%
% \institution{$^1$\{li,\hspace{0.1em} aizawa,\hspace{0.1em} matsui\}@hal.t.u-tokyo.ac.jp \hspace{0.3em} $^2$hinami@mantra.co.jp}
% \city{}
% \country{}}
% \renewcommand{\shortauthors}{author name and author name, et al.}

\begin{abstract}
    Recognizing characters and predicting speakers of dialogue are critical for comic processing tasks, such as voice generation or translation. However, because characters vary by comic title, supervised learning approaches like training character classifiers which require specific annotations for each comic title are infeasible. 
    This motivates us to propose a novel zero-shot approach, allowing machines to identify characters and predict speaker names based solely on unannotated comic images.
    In spite of their importance in real-world applications, these task have largely remained unexplored due to challenges in story comprehension and multimodal integration. 
    Recent large language models (LLMs) have shown great capability for text understanding and reasoning, while their application to multimodal content analysis is still an open problem. 
    To address this problem, we propose an iterative multimodal framework, the first to employ multimodal information for both character identification and speaker prediction tasks. 
    Our experiments demonstrate the effectiveness of the proposed framework, establishing a robust baseline for these tasks.
    Furthermore, since our method requires no training data or annotations, it can be used as-is on any comic series. 
\end{abstract}

%    <concept>
%        <concept_id>10010405.10010469.10010474</concept_id>
%        <concept_desc>Applied computing~Media arts</concept_desc>
%        <concept_significance>500</concept_significance>
%        </concept>
%  </ccs2012>

% \ccsdesc[500]{Applied computing~Media arts}


\begin{teaserfigure}
    \vspace{5mm}
    \includegraphics[width=\linewidth]{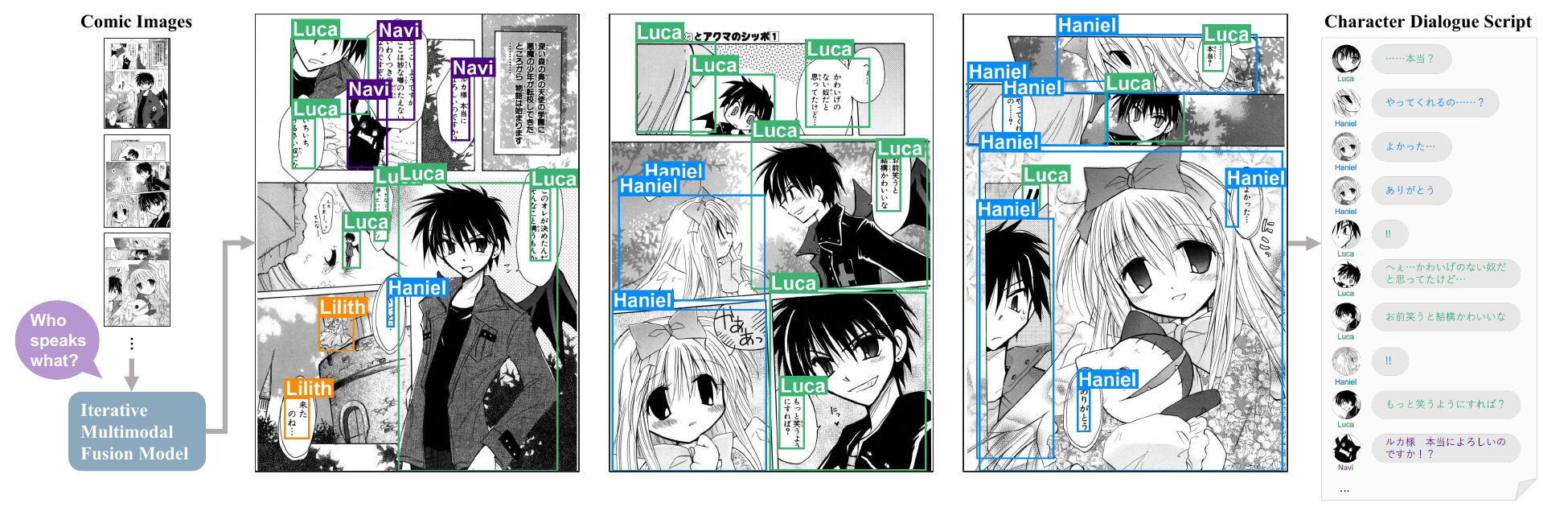}
    \caption{Our framework can predict character labels of unseen comics only from images. Courtesy of Kiriga Yuki.}
    \label{fig:introduction}
    \vspace{10mm}
\end{teaserfigure}


\maketitle

\section{Introduction}
The global comics market has experienced significant growth, sparking interest in computational analysis of comics to enrich user experience and accessibility.
Given that characters are central to comic storytelling, this paper focuses on two key tasks: \textit{character identification}, recognizing the characters in images, and \textit{speaker prediction}, predicting the speakers of specific dialogue.
These tasks enable diverse applications like voicing comics with character-specific voices and machine translations capturing each character's unique speech style.

Previous studies on speaker prediction in comics primarily addressed the correspondence between character regions and speech bubble regions in images~\cite{rigaud2015speech, li2024manga109dialog}, but they did not determine the speakers' names.
If we want to know the character names of the speakers, we first need to identify the character names of the character regions.
The straightforward supervised approach involves training models to detect and classify characters in images~\cite{ogawa2018object,zheng2020cartoon}. However, this approach requires annotations for each comic with different characters, which is impractical in the fast-paced comic industry that releases thousands of titles annually.

This motivates us to tackle the problem as a \textit{zero-shot} task: \textit{identifying characters and predicting speakers in new, unseen comics simply by analyzing the images, without need for any prior annotations}.

How does a human recognize speakers in new comics, without prior knowledge?
The process typically starts with noticing character names in the dialogue. 
For instance, if one character calls another ``Naruto", we associate the response to that dialogue with Naruto. We then connect the character's visual appearance with their dialogue.
When Naruto reappears in the following pages, even if his name is not mentioned in the dialogue, we can still know the character named Naruto.
Context from dialogues is also an important clue. Once we have determined the speaker of a particular dialogue, we can use contextual information to predict the speakers of nearby dialogues.
Additionally, this process allows us to update our knowledge of the characters' visual appearances across different pages.
After seeing many examples, we learn to identify characters and predict dialogue speakers even when their names are not directly mentioned in the text.

This process suggests two key challenges. 
(1) \textit{High-level text understanding}: To predict speakers from limited cues (e.g., names mentioned in dialogue), the system must interpret complex character interactions and story context throughout the whole book or chapter. 
(2) \textit{Multimodal integration}: To identify characters and predict the name of speakers without any annotations, integrating both visual and textual information is essential, thereby making a combination of text and image modalities indispensable.

Addressing these challenges, we propose a multimodal fusion approach. 
We leverage large language models (LLMs) for their context understanding and reasoning capabilities~\cite{zhang2023benchmarking, wei2022emergent, huang2022towards}, providing a strong baseline for zero-shot speaker prediction. 
To address the challenges of integrating LLMs with other modules and enhance the machine's comprehension of comics, we introduce an iterative framework. 
We merge text-based LLM predictions with image-based classifiers, and alternately refine each module using results from the other. 
This multimodal integration not only enables zero-shot character identification but also also notably improves text-only baseline in speaker prediction.
Moreover, by iteratively refining the integration of text and image information, this approach enhances the utilization of both modalities, thereby deepening the machine’s comprehension of comics.

Our contribution is twofold.
(1) \textbf{New tasks:} We are the first to integrate the tasks of character identification and speaker prediction in comics. 
Furthermore, our approach tackles zero-shot tasks without requiring any training or annotations, which are directly applicable to real-world scenarios.
(2) \textbf{Iterative multimodal fusion:} We pioneer in revealing the potential of LLMs for comics analysis and propose a novel method that integrates text and image modalities. 
To enhance the machine's deep understanding of comic content, we introduce an iterative framework aimed at progressively refining performance.
This is the first study to use both text and image information for character identification and speaker prediction, which are unexplored even outside zero-shot settings.

\section{Related Work}

\begin{figure*}[t]
    \centering
    \includegraphics[width=\linewidth]{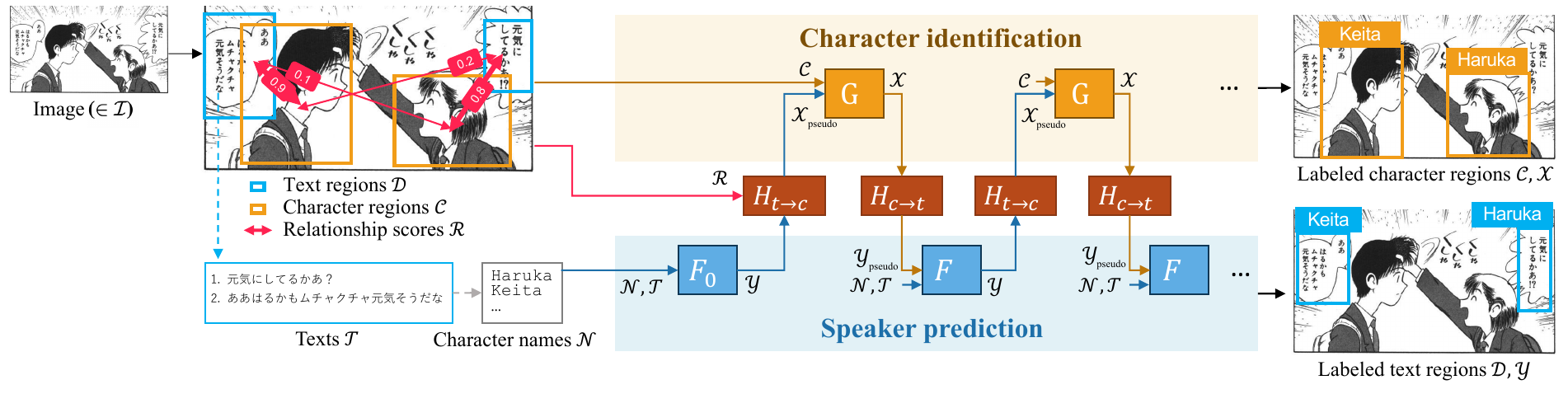}
    \caption{Our proposed framework for zero-shot character identification and speaker prediction in comics. Courtesy of Ito Shinpei.}
    \label{fig:pipeline}
\end{figure*}

Given the novelty of our study in predicting speaker names in comics, there are no existing studies directly related to this task. 
Therefore, we first explore related works from two distinct perspectives: comic speaker prediction and comic character identification.
Then, we introduce works related to our proposed method.

\subsection{Comic Speaker Prediction}
Previous studies of comic speaker prediction focus on predicting the correspondence between character bounding boxes and text regions.
Conventional methods are rule-based, primarily relying on the distance between the character and the text region~\cite{rigaud2015speech}.
They disregard visual and textual context information, causing them to fail in cases where the speaker is not closest to the text region. 
Recently, Li et al. employ scene graph generation (SGG) models~\cite{li2024manga109dialog} known for their effectiveness in visual relationship detection tasks~\cite{johnson2015image, xu2017scene, zellers2018neural, tang2020unbiased}, to predict the correspondence in a more robust way by using visual semantic information such as the face angle.
While the previous studies focused only on predicting the correspondence between character bounding boxes and text regions, we propose a new task to predict speaker identities, i.e., character names. This allows more applications such as text-to-speech with character-specific voices. Furthermore, while previous studies have utilized only visual information, we are the first to utilize both visual and textual information.

\subsection{Comic Character Identification}
Comic character identification is more challenging than face identification in real-world images due to the variance of drawing styles and poses.
Zheng et al. constructed a cartoon-face dataset and proposed an image classification model to classify characters~\cite{zheng2020cartoon}.
However, their supervised approach requires training character classifiers for each comic title, which limits the feasibility of this approach.
Some previous studies attempt unsupervised methods to handle unseen comics.
Tsubota et al. used deep features trained for generic face recognition in comics and adapted them to unseen comics~\cite{tsubota2018adaptation}.
Zhang et al. improved clustering performance on comics with a face-body combination and spatial-temporal correction~\cite{zhang2022unsupervised}.
However, they are limited to grouping the characters by clustering and cannot identify character names.
In addition, even though dialogue is an important cue to identify characters in comics, existing approaches have not used textual information.
In this paper, we propose the first multimodal approach that utilizes dialogue and predicts character names in unseen comics.

\subsection{Large Language Models}
The recent success of ChatGPT~\cite{chatgpt} and GPT-4~\cite{gpt4} has demonstrated the power of LLMs in understanding, generating, and interpreting human language with remarkable accuracy. Inspired by these advancements, we have pioneered the application of LLMs to the dialogue analysis in comics. 
Alongside the development of LLMs, large multimodal models (LMMs) such as LLaVA~\cite{liu2023llava, liu2023improved} and MiniGPT-4~\cite{zhu2023minigpt} have garnered attention for processing multimodal information.
% In spite of their capabilities in visual understanding and reasoning, currently, they can only handle a single or a few images as input. 
Although these models show strong performance on visual understanding and reasoning within a single panel, they struggle with understanding content across multiple panels~\cite{ikuta2024mangaubmangaunderstandingbenchmark}.
Comic analysis requires comprehension across longer contexts, such as multiple pages. Also, it involves learning character identity throughout the book or series. We thus propose an iterative framework to integrate multimodality and longer contexts into LLMs inference for this task.

\section{Approach}
\subsection{Problem Settings}
Let us define our problem setting for zero-shot character identification and speaker prediction.
The inputs are a sequence of page images of a specific comic $\mathcal{I}$.
The output consists of character regions $\mathcal{C} = \{c_i\}_{i=1}^N$, text regions $\mathcal{D} = \{d_j\}_{j=1}^M$,
and corresponding character labels $\mathcal{X} = \{x_i \in \mathcal{N}\}_{i=1}^N$ and $\mathcal{Y} = \{y_j \in \mathcal{N}\}_{j=1}^M$.
The character regions are the bodies of characters, and the text regions are speech bubbles.
$\mathcal{N}$ is the list of character names  (e.g., $\mathcal{N}= \{\texttt{Keitaro}, \texttt{Naru}, \ldots\}$), which is extracted from dialogues. These names serve as target labels for speaker prediction and character identification tasks in the following steps.
$N$ and $M$ denote the number of character and text regions in a given comic.

\noindentparagraph{\textbf{Data preprocessing.}}
Before initiating our main pipeline, we conducted a series of preprocessing steps.
First, character regions $\mathcal{C}$ and text regions $\mathcal{D}$ are obtained using object detectors for general comic character and text category, which are shown to achieve high accuracy~\cite{ogawa2018object}.
Subsequently, we derived the initial relationship scores $\mathcal{R}$ from $\mathcal{I}$, $\mathcal{C}$, and $\mathcal{D}$ using the SGG models~\cite{li2024manga109dialog}. 
Relationship scores $\mathcal{R} = \{r_{i \to j}\}_{i, j}$ represent the correspondences between character and text regions,
where $r_{i \to j} \in (0, 1)$ is the confidence that the character $c_i$ is the speaker of the text $d_j$.
This is used to propagate labels between the character and text regions.
Additionally, we extracted texts $\mathcal{T} = \{t_j\}_{j=1}^M$ from text regions $\mathcal{D}$ utilizing optical character recognition (OCR). 
Through text analysis by LLMs, we obtained the list of character names $\mathcal{N}$ from $\mathcal{T}$.

\begin{algorithm}[t]
\caption{Overall framework}
\label{alg:algorithm}
\begin{minipage}{\linewidth}
    \textbf{Input}: Images $\mathcal{I}$ \\
    \textbf{Parameter}: Iteration times $n$ \\
    \textbf{Intermediate Output}: Texts $\mathcal{T}$, Character names $\mathcal{N}$, Relationship scores $\mathcal{R}$ \\
    \textbf{Output}: Character regions $\mathcal{C}$, Text regions $\mathcal{D}$, Character labels $\mathcal{X}$, Text labels $\mathcal{Y}$
\end{minipage}
\begin{algorithmic}[1] %[1] enables line numbers
    \STATE Data preprocessing \\
    $\mathcal{C, D} \gets$ Object detection on $\mathcal{I}$ \\
    $\mathcal{R} \gets$ Initial relationship detection from $\mathcal{I, C, D}$ \\
    $\mathcal{T} \gets$ OCR extraction from $\mathcal{I,D}$ \\
    $\mathcal{N} \gets$ Character name extraction from $\mathcal{T}$
    \STATE Initial speaker prediction: Get initial labels of $\mathcal{Y}$ \\
    $\mathcal{Y} \gets$ $F_0(\mathcal{T, N})$
    \WHILE{Iteration times $\leq n$}
    \STATE Multimodal character identification: Update $\mathcal{X}$ \\
    $\mathcal{X}_\mathrm{pseudo} \gets$ $H_\mathrm{t \rightarrow c}(\mathcal{R}, \mathcal{Y})$ \\
    $\mathcal{X} \gets$ $G(\mathcal{C}, \mathcal{X}_\mathrm{pseudo})$ \\
    $\mathcal{R} \gets f_\mathrm{rescore} (\mathcal{R}, \mathcal{X}, \mathcal{Y})$ 
    \STATE Multimodal speaker prediction: Update $\mathcal{Y}$ \\
    $\mathcal{Y}_\mathrm{pseudo} \gets$ $H_\mathrm{c \rightarrow t}(\mathcal{R}, \mathcal{X})$ \\
    $\mathcal{Y} \gets$ $F(\mathcal{T, N}, \mathcal{Y}_\mathrm{pseudo})$ \\
    $\mathcal{R} \gets f_\mathrm{rescore} (\mathcal{R}, \mathcal{X}, \mathcal{Y})$ 
    \ENDWHILE
    \STATE \textbf{return} $\mathcal{C, D, X, Y}$
\end{algorithmic}
\end{algorithm}

\subsection{Overall Framework}
Our framework for zero-shot speaker prediction and character identification is illustrated in Figure~\ref{fig:pipeline}.
Speaker prediction and character identification are performed iteratively.
By using the output of each task as input for the other, we can exploit complementary multimodal information in both tasks, which leads to good performance even in zero-shot settings.

Our framework is comprised of three modules:
\begin{itemize}[leftmargin=15pt]
    \item \textbf{Speaker prediction:} Predict labels of text regions $\mathcal{Y}$ with LLMs. Initial predictions only use textual content, denoted as $F_0: (\mathcal{T,N}) \mapsto \mathcal{Y}$.
    From the second prediction, labels obtained from character identification $\mathcal{Y}_\mathrm{pseudo}$ is used, denoted as $F: (\mathcal{T, N}, \mathcal{Y}_\mathrm{pseudo}) \mapsto \mathcal{Y}$.
    \item \textbf{Character identification:} Predict labels of character regions $\mathcal{X}$ using image information and pseudo labels $\mathcal{X}_\mathrm{pseudo}$ obtained from speaker prediction.
    Denoted as $G: (\mathcal{C, X}_\mathrm{pseudo}) \mapsto \mathcal{X}$.
    \item \textbf{Label propagation:} Convert labels between character and text regions using relationship scores.
    Denoted as $H_\mathrm{t\rightarrow c}: (\mathcal{R, Y}) \mapsto \mathcal{X}$ and $H_\mathrm{c\rightarrow t}: (\mathcal{R, X}) \mapsto \mathcal{Y}$.
\end{itemize}

The procedure of our framework is shown in Algorithm~\ref{alg:algorithm}.
First, we predict speakers only from text information. 
This output is converted into labels of character regions using label propagation, whereby character identification is performed.
Then, we predict speakers again using the labels obtained in the previous step.
In addition, relationship scores are updated with $f_\mathrm{rescore}$ based on the predicted labels. 
These processes are repeated for a specific number of iterations.
In the following sections, we explain the details of each step.

\subsection{Initial Speaker Prediction}
\begin{figure}[t]
    \centering
    \includegraphics[width=1.0\linewidth]{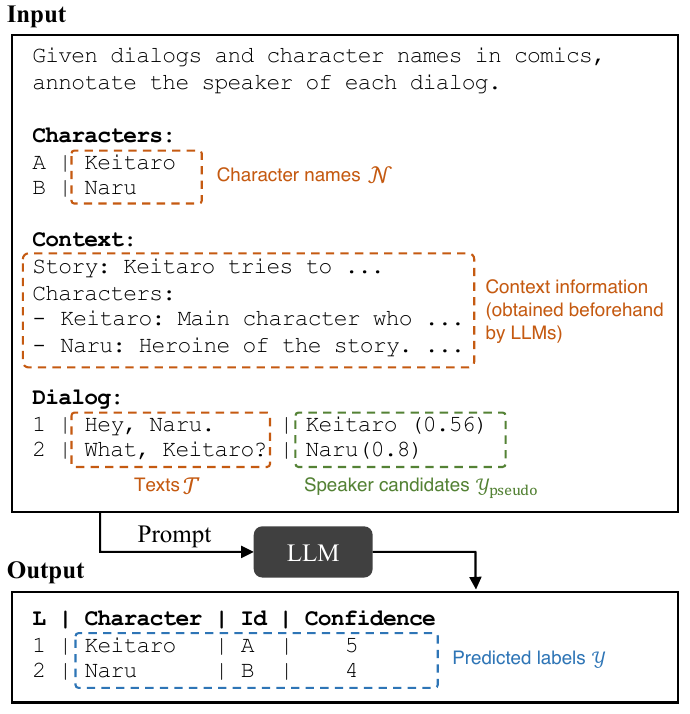}
    \caption{Speaker prediction with LLMs.}
    \label{fig:speaker_prediction}
\end{figure}

% We first predict speakers using only the textual content using LLMs as $F_0(\mathcal{T, N})$. GPT-4~\cite{gpt4} is used in this work.
We first use GPT-4~\cite{gpt4} to predict speakers only based on textual content, denoted as $F_0(\mathcal{T, N})$.
The input and outputs of LLMs are shown in Figure~\ref{fig:speaker_prediction}.
We input the dialogues $\mathcal{T}$ and the list of appearing characters' names $\mathcal{N}$ into the LLMs.
Due to output text length limitations of GPT-4, we split the dialogues into chunks.
To compensate for the missing context about the story in each chunk, we first extract context information about the story summary and character profile from $\mathcal{T}$ and $\mathcal{N}$ using LLMs.
This context is then fed into LLMs with each chunk as shown in Figure~\ref{fig:speaker_prediction}.
Note that speaker candidates $\mathcal{Y}_\mathrm{pseudo}$ are not input in the initial prediction.
As the output, we let LLMs output both character IDs and character names to get a stable output. These output character IDs are converted into $\mathcal{Y}$. 
In addition, we let LLMs output integer confidence scores from 1 to 5, and we exclude data with low confidence from subsequent steps.

\subsection{Multimodal Character Identification}
\begin{figure}[t]
    \centering
    \includegraphics[width=0.9\linewidth]{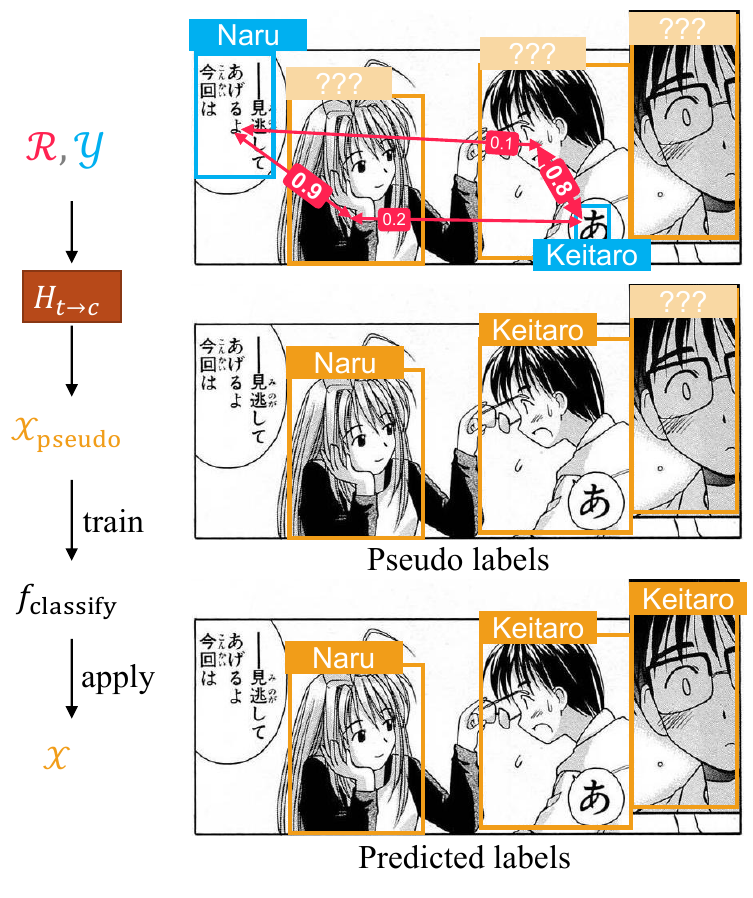}
    \caption{Pipeline of the character identification. Courtesy of Akamatsu Ken.}
    \label{fig:classifier}
\end{figure}

Given initial speaker prediction results $\mathcal{Y}$, we perform character identification as shown in Figure~\ref{fig:classifier}.

The first step is pseudo label generation with the label propagation module $H_\mathrm{t \rightarrow c}$.
For each character region, we select the character-text pair that has the highest relationship score among all combinations involving this character region. 
Then, we use the selected text region's label as the pseudo label of the character region.
To get higher precision, we set a threshold for the confidence of text region's label obtained by LLMs. In our experiments, removing data with confidence less than 3 lead to higher precision without reducing recall. 
In the second step, we construct a character classifier to identify the characters with unknown labels. 
We train an image classifier $f_\mathrm{classify}$ with pseudo labels $\mathcal{X}_\mathrm{pseudo}$ and then obtain $\mathcal{X}$ by applying $f_\mathrm{classify}$ to character regions $\mathcal{C}$.
We use ResNet50~\cite{he2016deep} as the image classifier.

As $\mathcal{Y}$ becomes closer to the true text labels, the generated pseudo labels will also be closer to the ground truth. 
Correspondingly, the training data for the character classifier becomes progressively more reliable, resulting in higher identification accuracy.
We iteratively enhance the performance of character identification by improving the precision of the speaker labels predicted by the speaker prediction module.

\subsection{Multimodal Speaker Prediction}
In the multimodal speaker prediction, we first generate pseudo labels for text regions $\mathcal{Y}_\mathrm{pseudo}$ from $\mathcal{X}$ by using the label propagation module $H_\mathrm{c \rightarrow t}$ in the same step as $H_\mathrm{t \rightarrow c}$.
% We treat these pseudo labels as candidate speakers for the texts, taking the confidence of character identification to be the confidence for the speaker candidates.
These pseudo labels serve as candidate speakers for the texts, with character identification confidence serving as the confidence for these candidates.
Confidences with less than 0.5 are filtered out from candidates.

Given $\mathcal{Y}_\mathrm{pseudo}$, the speaker is predicted by LLMs with the prompt in Figure~\ref{fig:speaker_prediction}.
We feed the LLMs both the confidence of character classifiers and the character name in a format such as \texttt{Keitaro (0.56)}, which informs the LLMs about the reliability of the given speaker candidates.

\subsection{Relationship Rescoring}
Relationship scores $\mathcal{R}$ are updated based on the predicted labels $\mathcal{X}$ and $\mathcal{Y}$.
Since character labels are not considered in the initial relationship scores, more accurate relationship scores can be obtained by considering character labels.
We use a simple rescoring method:
$r_{i,j}$ is multiplied by scale $s$ if $x_i=y_j$ and divided by $s$ if $x_i \neq y_j$. The scale $s$ is computed from confidences of labels $p_{x_i}$ and $p_{y_j}$ as $\min(1, \lambda p_{x_i} p_{y_j})$, where $\lambda$ is a hyper-parameter that is set to 2 in our experiments.
% With this step, pairs that are predicted as having the same character labels gets higher relationship scores, and vice versa.
With this step, pairs with the same character labels get higher relationship scores, and vice versa.
This is similar to the process where humans recognize characters and predict the speaker-text correspondences when reading comics.

\section{Experiments}
\begin{table*}[t]
\centering
\caption{Speaker prediction and character identification accuracy (\%).}
\label{tab:exp__main}
\begin{minipage}[c]{0.65\linewidth}
    \centering
    \begin{tabularx}{\textwidth}{@{}l*{9}{>{\centering\arraybackslash}X}}
    \toprule
    & & & & \multicolumn{3}{c}{Speaker pred.} & \multicolumn{3}{c}{Character id.} \\
    \cmidrule(l){5-7}\cmidrule(l){8-10}
    & iter & text & img  & \textit{Easy} & \textit{Hard} & \textit{Total} & \textit{Easy} & \textit{Hard} & \textit{Total} \\
    \midrule
    Baseline & & & & & & & & & \\
    ~~ K-means+Distance                           & - &           &\checkmark & 34.5\rlap{$^*$} & 31.8\rlap{$^*$} & 33.1\rlap{$^*$} & 37.0\rlap{$^*$} & 36.7\rlap{$^*$} & 36.8\rlap{$^*$} \\
    ~~ K-means+SGG                            & - &           &\checkmark & 36.7\rlap{$^*$} & 34.8\rlap{$^*$} & 35.7\rlap{$^*$} & 37.0\rlap{$^*$} & 36.7\rlap{$^*$} & 36.8\rlap{$^*$} \\
    \midrule Proposed & & & & & & & & & \\
    ~~ LLM only    & 0 &\checkmark &           & 41.8 & 45.1 & 43.6 & - & - & - \\
    ~~ Multimodal  & 1 &\checkmark &\checkmark & 51.0 & 51.2 & 51.1 & 45.8 & 39.6 & 42.4 \\
                                             & 2 &\checkmark &\checkmark & 52.4 & \textbf{51.3} & \textbf{51.8} & 48.5 & \textbf{40.3} & \textbf{44.0} \\
                                             & 3 &\checkmark &\checkmark & \textbf{53.5} & 49.8 & 51.6 & \textbf{48.9} & 37.7 & 42.8 \\
    \bottomrule
    \end{tabularx}
    \subcaption{Results on different test sets. $^*$ indicates that the baseline method used the ground truth to map clusters into labels, as explained in the experimental setup. }
\end{minipage}
\quad
\begin{minipage}[c]{0.32\linewidth}
    \centering
    \begin{tabularx}{\textwidth}{@{}l*{3}{>{\centering\arraybackslash}X}}
    \toprule
    & \small{iter} & \small{Speaker pred.} & \small{Character id.} \\
    \midrule Baseline & & & \\
    ~~ K-means+GT & - & 42.0\rlap{$^*$} & 36.8\rlap{$^*$} \\
    \midrule Proposed & & & \\
    ~~ LLM only    & 0 & 43.6 & - \\
    ~~ Multimodal  & 1 & 60.2 & 53.9 \\
                         & 2 & 63.4 & 55.5 \\
                         & 3 & \textbf{63.8} & \textbf{56.6} \\
    \bottomrule
    \end{tabularx}
    \subcaption{Results using the ground truth relationships.}
\end{minipage}
\end{table*}

\begin{figure*}[t]
    \centering
    \begin{minipage}[c]{0.44\linewidth}
        \centering
        \includegraphics[height=4.8cm]{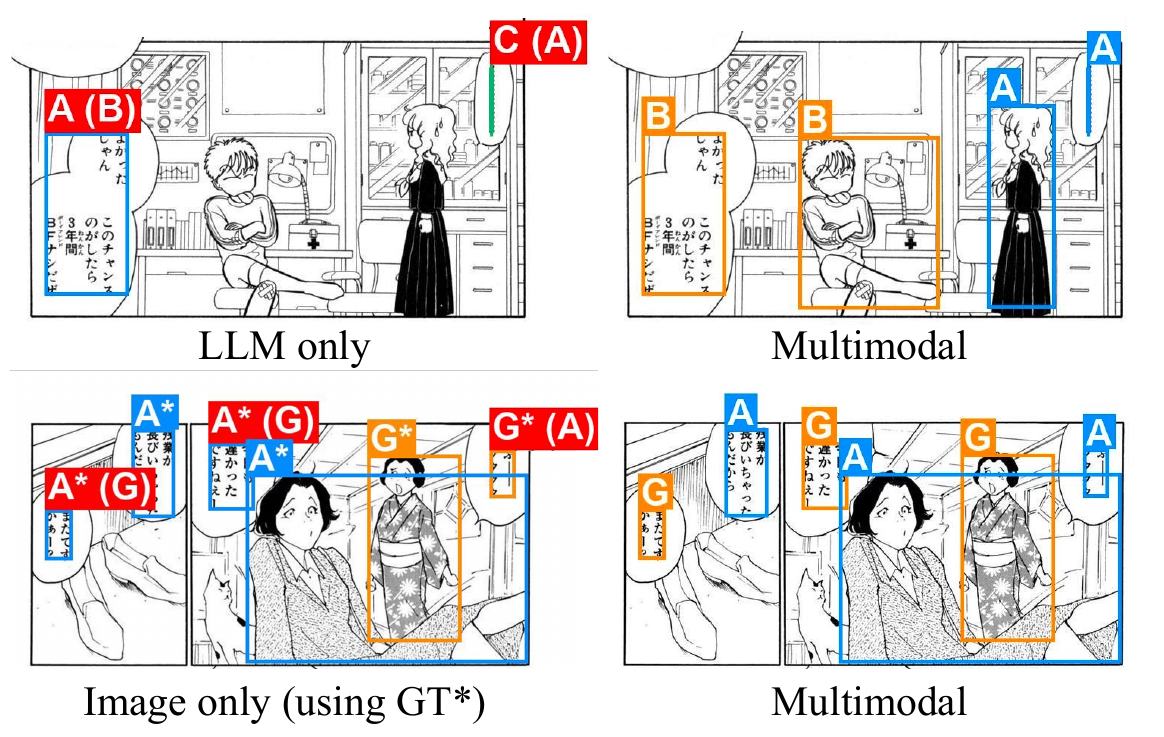}
        \subcaption{Unimodal vs. Multimodal}
    \end{minipage}
    \begin{minipage}[c]{0.55\linewidth}
        \centering
        \includegraphics[height=4.8cm]{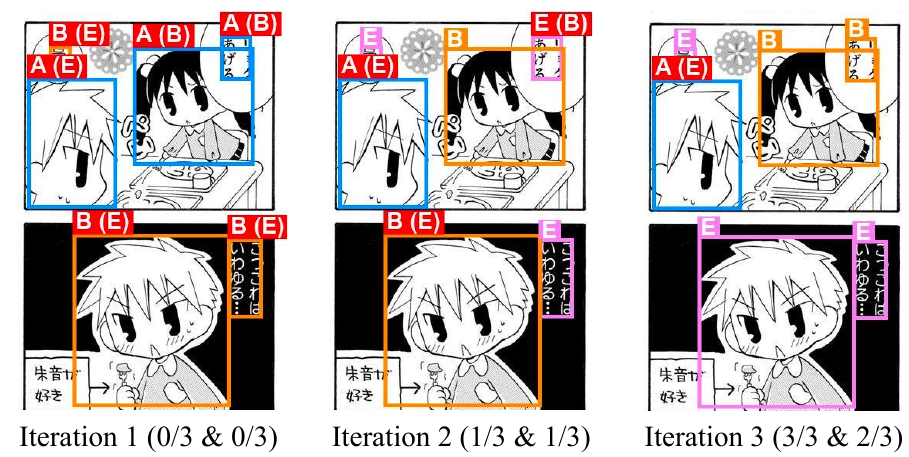}
        \subcaption{One-Step vs. Iterative (Accuracy of speaker pred. \& character id.)}
    \end{minipage}
    \caption{Examples of prediction results. Courtesy of Tashiro Kimu, Hikochi Sakuya, Tenya. Color of the bounding box indicates the predicted character label. Red labels indicate failure predictions. The labels on the boxes (e.g., `A') are character labels. Labels in brackets are the ground truth. (e.g., `A (B)' is the case where the ground truth is B but the prediction is A.)}
    \label{fig:exp__example}
\end{figure*}

\subsection{Experimental Setup}
\noindentparagraph{\textbf{Dataset.}}
We used the Manga109 dataset~\cite{aizawa2020building, ogawa2018object}, which comprises 109 volumes of Japanese comics and provides character labels for both character and text regions. 
For the zero-shot setting, we constructed the training and testing sets from distinct comic titles; the characters in the testing set were unseen in the training set.
We selected 23 volumes as the testing set.
The remaining volumes were used for training and validation of the relationship detection module and pre-training of the character classifier.

For the character regions, we used the body region annotations~\cite{ogawa2018object}, which represent the entire bodies of the characters. 
We preferred this method because whole-body regions offer more information (such as clothing and body types) for character identification than face regions.
Besides, comic datasets often include characters with body-only annotations, such as those shown from the back or non-human characters, which necessitates the use of body regions for identification.
For the character labels associated with the text regions, we used Manga109Dialog annotations~\cite{li2024manga109dialog}.

\noindentparagraph{\textbf{Task Settings.}}
For each comic in the test set of Manga109, 
we used page images $\mathcal{I}$ and a list of character names as inputs. 
Since object detection and OCR are not the main focuses of this paper, and since the results of character name extraction can significantly impact the accurate evaluation of our main pipeline's performance, 
in the main evaluation, we omit the steps of obtaining $\mathcal{C, D, T}$, and $\mathcal{N}$ in data preprocessing. 
Instead, we regard them as known information by utilizing the annotations of Manga109.
This simplifies the evaluation and makes the experiment replication easier while maintaining the main challenges of our tasks.
End-to-end zero-shot settings were evaluated in the final experiments.
Given that some titles in Manga109 feature a large number of characters, we excluded characters with an appearance frequency of less than 3\% from the list of character names in our experiments. However, to ensure a fair comparison with future works, we still predicted and evaluated the labels of these infrequently appearing characters.

% \paragraph{Evaluation.}
Under the above settings, the tasks of speaker prediction and character identification were classifying the character labels for the character and text regions ($\mathcal{X}$ and $\mathcal{Y}$). 
We calculated the accuracy of classification results to evaluate these tasks, which is the ratio of correctly predicted regions to total regions.
Additionally, we calculated precision and recall for the pseudo labels in the ablation study. 
Precision is the ratio of correct pseudo labels to the total number of generated pseudo labels. 
Recall is the ratio of correct pseudo labels to the total number of regions.

\noindentparagraph{\textbf{Relationship Prediction.}}
Our method used the relationship scores $\mathcal{R}$ to show correspondences between character and text regions. 
We compared three types of initial relationship scores: \textbf{SGG:} a deep learning-based relationship prediction using the scene graph generation (SGG) model~\cite{li2024manga109dialog}, 
\textbf{Distance:} a rule-based method using the distance between the center coordinates of the character and text regions~\cite{rigaud2015speech}, and \textbf{GT:} ground truth annotations of Manga109Dialog. 
We used SGG in the main experiments and used Distance and GT in the ablation study. 
The confidence of SGG model was used as the score for SGG, and the reciprocal of the distance was used as the score for Distance.
In the case of GT, we took the relationship score to be 1.0 for all pairs.

\noindentparagraph{\textbf{Speaker Prediction with LLMs.}}
We used the GPT-4 model~\cite{gpt4} (specifically, \texttt{gpt4-0314}) for speaker prediction. 
First, we produced a context summary by feeding all dialogue to GPT-4. 
Next, we performed speaker prediction by inputting both the extracted context information and the dialogue text itself into the model. 
Due to GPT-4's output token limitation, we divided every conversation into segments, each comprising 60 sentences. 
A complete list of these prompts is available in the supplementary material.

\noindentparagraph{\textbf{Character Classifier Training.}} 
We used ResNet50~\cite{he2016deep} as the character region classifier, initially pre-trained on ImageNet~\cite{deng2009imagenet}. 
We chose this model as the classifier due to its robustness in handling the diverse and complex visual patterns of comic character bodies, which provides more advantages over simply fine-tuning face recognition models.
We also explored state-of-the-art classification models like ConvNeXt\cite{liu2022convnet} and models pre-trained on the anime dataset Danbooru\cite{danbooru2021}. 
However, our experiments demonstrated that the fine-tuned ResNet model offers the best performance. 

Our training process involved two steps: pre-training for generic comics and fine-tuning for individual unseen comic in the test set.
Pre-training, done only once for generic comics, aims at domain adaptation from the ImageNet pre-trained model to comic characters.
Manga109 training set with ground truth annotations of 349 characters is used for pre-training.
We fine-tune on each individual unseen comic using pseudo labels generated through our multimodal iterative fusion process.
We employ various techniques, including data augmentation and model ensembling, to achieve stable results from training with noisy labels. 
Since this module itself is not the main focus of this paper, we describe the details in the supplementary material.

\noindentparagraph{\textbf{Baselines.}}
Since no existing method can predict character labels in a zero-shot setup, we constructed our own baselines.
For character identification task, we first group character regions of each comic by the clustering of deep features obtained using the Manga109 pre-trained model explained above. K-means clustering~\cite{lloyd1982least} with $k$=$\lvert \mathcal{N} \rvert$ is used.
Since it is impossible to assign character labels to each cluster without text information, we map each cluster to the character labels using ground truth so that accuracy is maximized, which can be regarded as an upper bound of the clustering approach. 
The labels of character regions are converted into those of text regions using relationship prediction methods~\cite{rigaud2015speech, li2024manga109dialog}, which are
referred to as \textbf{K-means + Distance} and \textbf{K-means + SGG}, respectively in Table~\ref{tab:exp__main}. 

\subsection{Main Results}
Table~\ref{tab:exp__main} shows the results of the proposed method and the baselines. 
The \textit{text} and \textit{img} columns indicate the used modalities for each method. 
Each iteration involves identifying characters based on the speaker predictions from the previous iteration, followed by a new round of speaker prediction that uses the updated character labels and relationship scores. 
The initial phase, where the speaker prediction is conducted only with textual information via LLMs, is iteration 0.
The subsequent complete iteration cycle following iteration 0 is iteration 1.
To validate the effectiveness of our proposed method, we divided the test set (\textit{Total}) into \textit{Easy} and \textit{Hard} by the difficulty of relationship prediction. 
\textit{Easy} contains 11 volumes with an accuracy of relationship prediction over 75\%. 
The remaining 12 volumes were categorized as \textit{Hard}. 

As shown in Table~\ref{tab:exp__main} (a), our proposed multimodal approach produces a significant improvement in accuracy over unimodal methods in both speaker prediction and character identification.
For speaker prediction, the accuracy of \textit{LLM only} was 43.6\% but it increased to 51.1\% in the first iteration. 
The accuracy of character identification also reached 42.4\%, outperforming the baseline that uses the ground truth labels.
Although the accuracy of character identification was less than 50\%, it is noteworthy that our method is the first one that recognizes character labels in unseen comics.
The difficulty of identifying less frequent characters in zero-shot settings causes the low accuracy in both tasks, which is posed as future work.

Across all test sets, the results from iteration 2 showed an improvement over iteration 1. 
In iteration 3, while the accuracy for \textit{Easy} data continued to increase, there was a slight decline observed for \textit{Hard} and \textit{Total} data. 
This suggests that there are limits to the accuracy gains in relationship detection from further iterations.
To investigate this further, we do an analysis using the ground truth relationships in Table~\ref{tab:exp__main} (b), where the accuracy of both speaker prediction and character identification keeps increasing with further iterations.
These results suggest our iterative process is more effective in the case that the prediction of relationships between text and character regions is accurate.

\begin{figure}[t]
    \centering
    \includegraphics[width=1.0\linewidth]{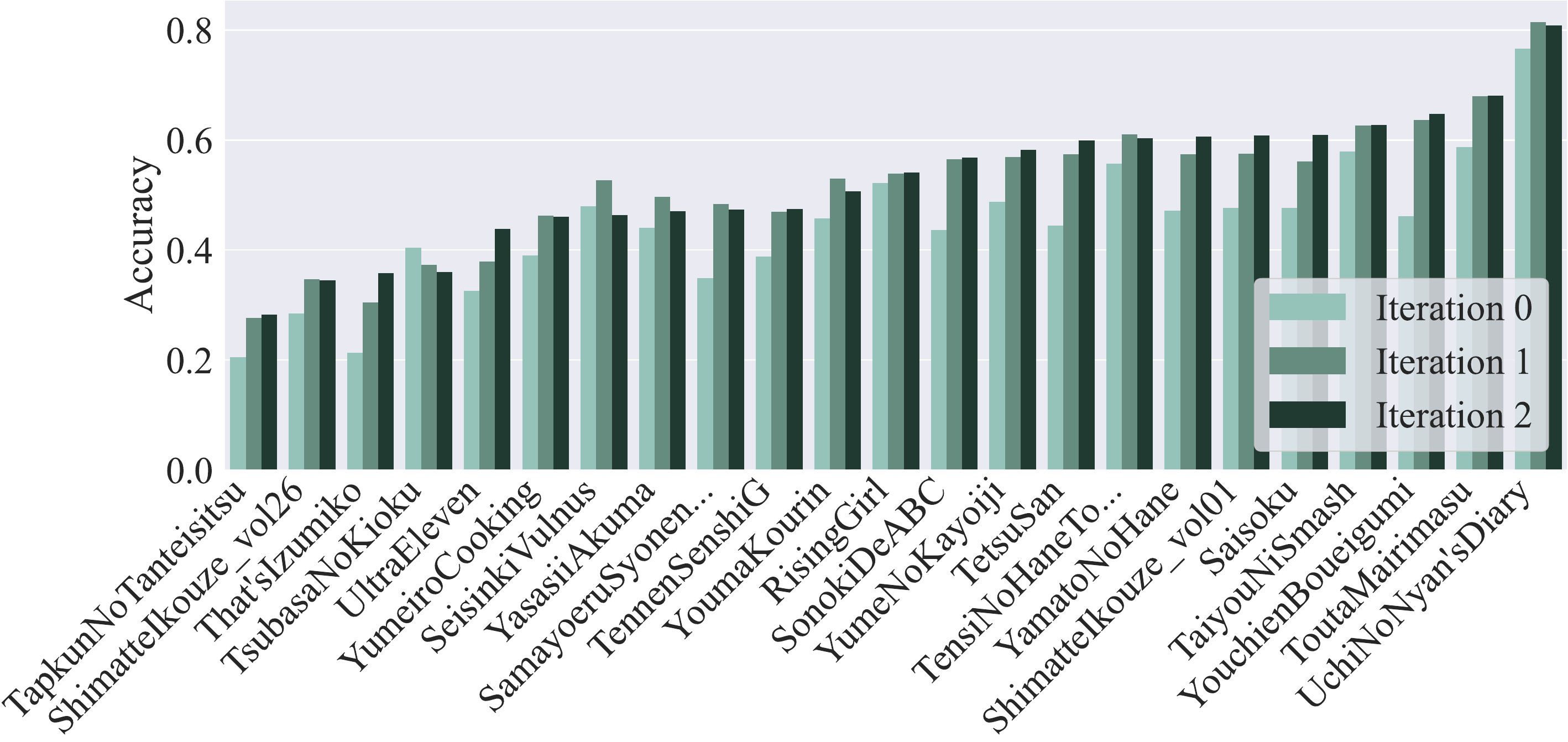}
    \caption{Speaker prediction accuracy of each comic title.}
    \label{fig:exp__per_manga}
\end{figure}

\begin{figure}[t]
    \centering
    \includegraphics[width=1.0\linewidth]{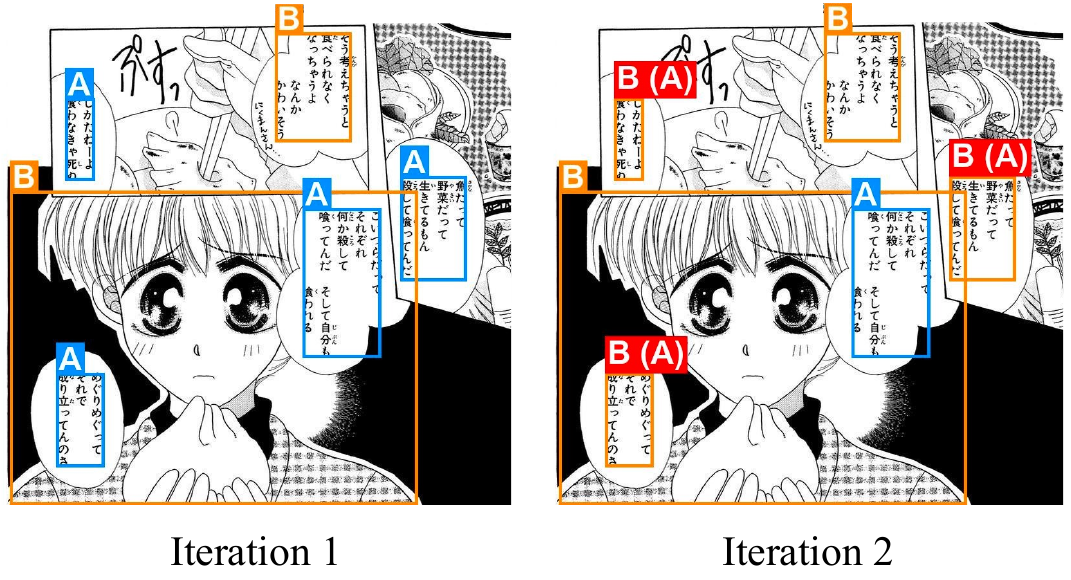}
    \caption{Failure cases. Courtesy of Hanada Sakumi.}
    \label{fig:exp__difficult}
\end{figure}

Figure~\ref{fig:exp__example} shows examples of prediction results. 
We evaluate the effectiveness of the proposed method in two aspects: \textbf{Unimodal vs. Multimodal} (Figure~\ref{fig:exp__example} (a)) and \textbf{One-Step vs. Iterative} (Figure~\ref{fig:exp__example} (b)).
Given that unimodal approaches are incapable of achieving zero-shot character identification, when using only LLM, character identification is not performed. 
\textit{Image only} results corresponds to the baseline k-means+SGG: labels of character regions are derived from ground truth.
Speaker predictions from texts alone present challenges for LLMs, especially with texts lacking distinctive character features, while it is easy to make correct predictions when integrating visual information.
Conversely, when speaker predictions are made only based on images, such as by selecting the character closest to the text as the speaker, the method fails in two scenarios:
(1) when the speaker is not the character closest to the text;
(2) the speaker does not appear in the image.
In contrast, our multimodal approach handles these complexities well.
Figure~\ref{fig:exp__example} (b) shows the results of the proposed method in each iteration. 
As the number of iterations increases, there is a clear trend of increasing accuracy in both character identification and speaker prediction tasks.
Although the first iteration yields no correct predictions, by the third iteration, the accuracy has significantly improved.
These findings affirm that our method can improve the accuracy of both tasks by iteratively refining the results using multimodal information.

Figure~\ref{fig:exp__per_manga} shows the accuracy of speaker prediction for each comic. 
The three bars in each comic indicate the accuracy of iteration 0 (LLM only), iteration 1, and iteration 2, respectively.
The iterations improved accuracy on most of the comics, which shows the effectiveness of our iterative approach on different comics. 
This figure also shows that performance can vary significantly across different comics, ranging from 0.2 to 0.8. 
Note that more iterations do not necessarily lead to higher accuracy. This is because if the accuracy of relationship prediction or character identification is low, the candidates can introduce noise into speaker prediction. 
An example is shown in Figure~\ref{fig:exp__difficult}. 
In iteration 1, the LLMs correctly predicted the speaker (character A). However, due to the distance between the speaker and the text, character A received a low relationship score. 
In contrast, character B, being closer and having a higher relationship score, was incorrectly identified as the speaker candidate with high confidence. 
This led to an incorrect prediction in iteration 2, which explains why there is a decrease in accuracy by iterations in Figure~\ref{fig:exp__per_manga} and Table~\ref{tab:exp__main}.

% ----------------------------------------------------
% Ablation: Speaker candidate
% ----------------------------------------------------
\subsection{Ablation Study}

\begin{table}[t]
\centering
\caption{Precision and recall of speaker candidates and the speaker prediction accuracy examined by varying the relationship scores.}
\label{tab:exp__speaker_candidate}
\begin{tabular}{@{}lcccc@{}}
\toprule
& \multicolumn{3}{c}{Speaker candidates} & Speaker pred. \\
\cmidrule(lr){2-4}\cmidrule(l){5-5}
Relation & Prec & Recall & Accuracy & Accuracy \\
\midrule
         Distance & 60.7 & 29.0 & 44.6 & 48.5 \\
         SGG      & 60.8 & 32.4 & 45.1 & 51.1 \\
         GT       & 74.0 & 39.5 & 54.6 & 60.2 \\
\bottomrule
\end{tabular}
\end{table}

\begin{table}[t]
\centering
\caption{Precision and recall of pseudo labels and impact on character identification accuracy.}
\label{tab:exp__pseudo_label}
\begin{tabular}{@{}lccccc@{}}
\toprule
& \multicolumn{3}{c}{Pseudo labels} & Character id. \\
\cmidrule(lr){2-4}\cmidrule(l){5-5}
Relation & Prec & Recall & Accuracy & Accuracy \\
\midrule 
Distance & 32.6 & 29.1 & 29.9 & 41.6 \\
SGG      & 34.2 & 28.4 & 29.2 & 42.4 \\
GT       & 58.8 & 23.5 & 25.1 & 53.9 \\
\bottomrule
\end{tabular}
\end{table}

\noindentparagraph{\textbf{Pseudo Labels.}}
Table~\ref{tab:exp__speaker_candidate} and \ref{tab:exp__pseudo_label} shows the quality of the pseudo labels $\mathcal{X}_\mathrm{pseudo}$ and speaker candidates $\mathcal{Y}_\mathrm{pseudo}$ and their effect on accuracy for speaker prediction and character identification tasks. Three relationship scores are compared. 2nd and 3rd columns are the precision and recall of the pseudo labels. The accuracies listed in the 4th column are those of the pseudo labels before thresholding. These are shown for comparison with the final performance score (5th column). 
The carryover from pseudo labels’ accuracies (4th column) to final accuracy (5th column) explains the effectiveness of our iterative process (+6.0\% and +13.2\% gains in speaker prediction and character identification, respectively when using SGG). It shows that the result obtained from the previous step is refined in each module by using different modal information.
By comparing relationship scores, we can see the result gets better by improving the quality of pseudo labels with better relationship prediction; specifically, the precision of GT in Table~\ref{tab:exp__pseudo_label} is 24.6\% higher than that of SGG, which leads to 10.5\% higher accuracy
(recall and accuracy with GT is low because non-speaking character regions are not associated with text regions.)
This suggests the higher precision in pseudo labels affects the character identification performance a lot.
In our current method, the combination of the noises of speaker and relationship prediction decreases the precision.
Getting reliable pseudo labels is posed as future work.

% ----------------------------------------------------
% Ablation: Propmts
% ----------------------------------------------------
\noindentparagraph{\textbf{LLM Prompts.}}

\begin{table}[t]
\centering
\caption{Speaker prediction accuracy with different prompts.}
\label{tab:exp__prompt}
\begin{tabular}{@{}lccccc@{}}
\toprule
& iter & ctx & cand & prob & Speaker pred. \\
\midrule
LLM only   & 0 &            &            &            & 38.9 \\
           & 0 & \checkmark &            &            & 43.6 \\
\midrule
Multimodal & 1 & \checkmark & \checkmark &            & 49.1 \\
           & 1 & \checkmark & \checkmark & \checkmark & \textbf{51.1} \\
\bottomrule
\end{tabular}
\end{table}

\begin{table}[t]
\centering
\caption{Relationship prediction and speaker prediction accuracy. SGG w/ rescoring is the result after the first iteration.}
\label{tab:exp__relation}
\begin{tabular}{@{}lcc@{}}
\toprule
% \cmidrule(lr){2-2}\cmidrule(l){3-3}
% & Accuracy$_\mathrm{t \rightarrow c}$  & Accuracy \\
& Relation pred. & Speaker pred. \\
\midrule
Distance                       & 71.9          & 47.7 \\
SGG~(Li et al. 2023)           & 78.1          & 50.3 \\
SGG w/ rescoring      & \textbf{79.8} & \textbf{51.1} \\
\bottomrule
\end{tabular}
\end{table}

Table~\ref{tab:exp__prompt} shows the results for different LLM prompts in the speaker prediction. The three options in the table correspond to 
(1) \textit{ctx}: context information about characters and stories, 
(2) \textit{cand}: speaker candidates, 
and (3) \textit{prob}: the probability of candidates.
Their respective improvements in accuracy are 4.7\%, 5.5\%, and 2.0\%.
Each option introduces information not present in the dialogues of each chunk: context provides whole-story information, 
while the speaker candidates and their probabilities are obtained from the images. 
These options offer complementary information to each other. The combination of all three options, as shown in Figure~\ref{fig:speaker_prediction}, performed best. 

% ----------------------------------------------------
% Ablation: Relationship score
% ----------------------------------------------------
\noindentparagraph{\textbf{Relationship Rescoring.}}
Table~\ref{tab:exp__relation} shows the effect of the proposed relationship rescoring
on relationship prediction and speaker prediction.
We evaluated the relationship prediction in the manner described in the previous work~\cite{li2024manga109dialog}:
for each text region, the corresponding character region with the highest relationship score was selected. 
Then, the accuracy was calculated by dividing the number of correctly matched regions by the total number of text regions. 
Our approach with rescoring achieved an accuracy 79.8\%, which is better than previous methods.
Moreover, it increased speaker prediction accuracy. 
In contrast to the previous work that doesn't consider character labels, our method can predict the relationships better by predicting character labels of both images and texts.

% ----------------------------------------------------
% Zero-shot examples
% ----------------------------------------------------
\subsection{Zero-shot Evaluation}
We evaluated our method in entirely zero-shot settings using only input images. 
The bounding boxes of the character and the text regions were detected with Faster R-CNN~\cite{ren2015faster}, and dialogues were extracted by using OCR in the manner described in~\cite{hinami2021towards}.
We extracted a list of character names directly from the dialogues using GPT-4.
These names then served as target labels for speaker prediction and character identification tasks.

Table~\ref{tab:exp__zero_shot} shows the experimental results under entirely zero-shot settings.
In assessing the experimental results, we manually mapped extracted names to true names.
Each predicted region is counted as correct if it was detected with an IoU \textgreater 0.5 and was correctly labeled.
Our method achieved 38.7\% and 35.0\% in accuracy for speaker prediction and character identification tasks, respectively.
This is lower than the results shown in Table~\ref{tab:exp__main} due to errors in object detection, OCR, and character name extraction. 
Our experiments demonstrated that the impact of errors from object detection and OCR on overall performance is limited.
When provided only with images and a list of character names, we achieved 45.9\% and 40.0\% in accuracy, respectively. 
To show the effect of character name extraction on the performance, we computed the \textit{upper bound} in Table~\ref{tab:exp__zero_shot}, which is the accuracy under ideal conditions, where all labels of extracted names are perfectly predicted. 
The recall for extracted names was 62.7\%, thereby limiting the upper bounds to 67.3\% for speaker prediction and 63.9\% for character identification. 
We analyzed the accuracy across different comic titles in an entirely zero-shot setting and found significant variations. 
These variations arise because character names are rarely mentioned in the dialogues of several titles, posing a challenge that is difficult even for humans. 
However, we observed promising results in titles where character names appear frequently in dialogues and the relationships between characters and texts are clear, as examples shown in Figure~\ref{fig:introduction}.

\begin{table}[t]
\centering
\caption{Accuracy in an entirely zero-shot setting.}
\label{tab:exp__zero_shot}
\begin{tabular}{@{}lccc@{}}
\toprule
& iter & Speaker pred. & Character id. \\
\midrule
LLM only    & 0 & 34.1 & - \\
\midrule
Multimodal  & 1 & 37.7 & \textbf{35.6} \\
            & 2 & \textbf{38.7} & 35.0 \\
            % & 3 & 37.9 & 33.8 \\
\midrule
Upper bound & - & 67.3 & 63.9 \\
\bottomrule
\end{tabular}
\end{table}

\section{Conclusion}
We are the first to introduce zero-shot character identification and speaker prediction in comics.
To solve this unexplored and challenging task, our proposed framework utilizes multimodal integration and high-level context understanding. Since story understanding and multimodal integration are key in comics analysis, our promising results can be a key milestone that opens new directions in comic analysis.
In addition, our method can be applied to any comics immediately without the need to train on specific comics, bridging the gap to real-world application.
Moreover, our framework can be easily adapted to different domains because each module of our framework is modular and reusable (e.g., learning person classifiers on video data).
Since the potential of multimodal speaker prediction extends beyond comics, ranging from movies to online conferences, we hope that our work will impact other fields as well.

% \begin{acks}
% \end{acks}

\bibliographystyle{ACM-Reference-Format}
\balance
\bibliography{sample-base}


\begin{thebibliography}{31}

%%% the Web address.

\ifx \showCODEN    \undefined \def \showCODEN     #1{\unskip}     \fi
\ifx \showDOI      \undefined \def \showDOI       #1{#1}\fi
\ifx \showISBNx    \undefined \def \showISBNx     #1{\unskip}     \fi
\ifx \showISBNxiii \undefined \def \showISBNxiii  #1{\unskip}     \fi
\ifx \showISSN     \undefined \def \showISSN      #1{\unskip}     \fi
\ifx \showLCCN     \undefined \def \showLCCN      #1{\unskip}     \fi
\ifx \shownote     \undefined \def \shownote      #1{#1}          \fi
\ifx \showarticletitle \undefined \def \showarticletitle #1{#1}   \fi
\ifx \showURL      \undefined \def \showURL       {\relax}        \fi
% invisible to TeX
\providecommand\bibfield[2]{#2}
\providecommand\bibinfo[2]{#2}
\providecommand\natexlab[1]{#1}
\providecommand\showeprint[2][]{arXiv:#2}

\bibitem[Aizawa et~al\mbox{.}(2020)]%
        {aizawa2020building}
\bibfield{author}{\bibinfo{person}{Kiyoharu Aizawa}, \bibinfo{person}{Azuma Fujimoto}, \bibinfo{person}{Atsushi Otsubo}, \bibinfo{person}{Toru Ogawa}, \bibinfo{person}{Yusuke Matsui}, \bibinfo{person}{Koki Tsubota}, {and} \bibinfo{person}{Hikaru Ikuta}.} \bibinfo{year}{2020}\natexlab{}.
\newblock \showarticletitle{Building a manga dataset “manga109” with annotations for multimedia applications}.
\newblock \bibinfo{journal}{\emph{IEEE MultiMedia}} \bibinfo{volume}{27}, \bibinfo{number}{2} (\bibinfo{year}{2020}), \bibinfo{pages}{8--18}.
\newblock


\bibitem[Anonymous et~al\mbox{.}(2022)]%
        {danbooru2021}
\bibfield{author}{\bibinfo{person}{Anonymous}, \bibinfo{person}{Danbooru community}, {and} \bibinfo{person}{Gwern Branwen}.} \bibinfo{year}{2022}\natexlab{}.
\newblock \bibinfo{title}{Danbooru2021: A Large-Scale Crowdsourced \& Tagged Anime Illustration Dataset}.
\newblock \bibinfo{howpublished}{\url{https://gwern.net/danbooru2021}}.
\newblock


\bibitem[Deng et~al\mbox{.}(2009)]%
        {deng2009imagenet}
\bibfield{author}{\bibinfo{person}{Jia Deng}, \bibinfo{person}{Wei Dong}, \bibinfo{person}{Richard Socher}, \bibinfo{person}{Li-Jia Li}, \bibinfo{person}{Kai Li}, {and} \bibinfo{person}{Li Fei-Fei}.} \bibinfo{year}{2009}\natexlab{}.
\newblock \showarticletitle{Imagenet: A large-scale hierarchical image database}. In \bibinfo{booktitle}{\emph{Proceedings of the IEEE conference on computer vision and pattern recognition}}. \bibinfo{pages}{248--255}.
\newblock


\bibitem[He et~al\mbox{.}(2016)]%
        {he2016deep}
\bibfield{author}{\bibinfo{person}{Kaiming He}, \bibinfo{person}{Xiangyu Zhang}, \bibinfo{person}{Shaoqing Ren}, {and} \bibinfo{person}{Jian Sun}.} \bibinfo{year}{2016}\natexlab{}.
\newblock \showarticletitle{Deep residual learning for image recognition}. In \bibinfo{booktitle}{\emph{Proceedings of the IEEE conference on computer vision and pattern recognition}}. \bibinfo{pages}{770--778}.
\newblock


\bibitem[Hinami et~al\mbox{.}(2021)]%
        {hinami2021towards}
\bibfield{author}{\bibinfo{person}{Ryota Hinami}, \bibinfo{person}{Shonosuke Ishiwatari}, \bibinfo{person}{Kazuhiko Yasuda}, {and} \bibinfo{person}{Yusuke Matsui}.} \bibinfo{year}{2021}\natexlab{}.
\newblock \showarticletitle{Towards fully automated manga translation}. In \bibinfo{booktitle}{\emph{Proceedings of the AAAI conference on artificial intelligence}}, Vol.~\bibinfo{volume}{35}. \bibinfo{pages}{12998--13008}.
\newblock


\bibitem[Huang and Chang(2023)]%
        {huang2022towards}
\bibfield{author}{\bibinfo{person}{Jie Huang} {and} \bibinfo{person}{Kevin Chen-Chuan Chang}.} \bibinfo{year}{2023}\natexlab{}.
\newblock \showarticletitle{Towards reasoning in large language models: A survey}. In \bibinfo{booktitle}{\emph{Proceedings of the 61th annual meeting of the association for computational linguistics}}. \bibinfo{pages}{1049--1065}.
\newblock


\bibitem[Ikuta et~al\mbox{.}(2024)]%
        {ikuta2024mangaubmangaunderstandingbenchmark}
\bibfield{author}{\bibinfo{person}{Hikaru Ikuta}, \bibinfo{person}{Leslie Wöhler}, {and} \bibinfo{person}{Kiyoharu Aizawa}.} \bibinfo{year}{2024}\natexlab{}.
\newblock \bibinfo{title}{MangaUB: A Manga Understanding Benchmark for Large Multimodal Models}.
\newblock
\newblock
\showeprint[arxiv]{2407.19034}


\bibitem[Johnson et~al\mbox{.}(2015)]%
        {johnson2015image}
\bibfield{author}{\bibinfo{person}{Justin Johnson}, \bibinfo{person}{Ranjay Krishna}, \bibinfo{person}{Michael Stark}, \bibinfo{person}{Li-Jia Li}, \bibinfo{person}{David Shamma}, \bibinfo{person}{Michael Bernstein}, {and} \bibinfo{person}{Li Fei-Fei}.} \bibinfo{year}{2015}\natexlab{}.
\newblock \showarticletitle{Image retrieval using scene graphs}. In \bibinfo{booktitle}{\emph{Proceedings of the IEEE conference on computer vision and pattern recognition}}. \bibinfo{pages}{3668--3678}.
\newblock


\bibitem[Krizhevsky et~al\mbox{.}(2012)]%
        {krizhevsky2012imagenet}
\bibfield{author}{\bibinfo{person}{Alex Krizhevsky}, \bibinfo{person}{Ilya Sutskever}, {and} \bibinfo{person}{Geoffrey~E Hinton}.} \bibinfo{year}{2012}\natexlab{}.
\newblock \showarticletitle{Imagenet classification with deep convolutional neural networks}. In \bibinfo{booktitle}{\emph{Advances in neural information processing systems}}, Vol.~\bibinfo{volume}{25}.
\newblock


\bibitem[Li et~al\mbox{.}(2024)]%
        {li2024manga109dialog}
\bibfield{author}{\bibinfo{person}{Yingxuan Li}, \bibinfo{person}{Kiyoharu Aizawa}, {and} \bibinfo{person}{Yusuke Matsui}.} \bibinfo{year}{2024}\natexlab{}.
\newblock \showarticletitle{Manga109Dialog: A Large-scale Dialogue Dataset for Comics Speaker Detection}. In \bibinfo{booktitle}{\emph{Proceedings of the IEEE International Conference on Multimedia and Expo}}.
\newblock


\bibitem[Liu et~al\mbox{.}(2023a)]%
        {liu2023improved}
\bibfield{author}{\bibinfo{person}{Haotian Liu}, \bibinfo{person}{Chunyuan Li}, \bibinfo{person}{Yuheng Li}, {and} \bibinfo{person}{Yong~Jae Lee}.} \bibinfo{year}{2023}\natexlab{a}.
\newblock \bibinfo{title}{Improved baselines with visual instruction tuning}.
\newblock
\newblock
\showeprint[arxiv]{2310.03744}


\bibitem[Liu et~al\mbox{.}(2023b)]%
        {liu2023llava}
\bibfield{author}{\bibinfo{person}{Haotian Liu}, \bibinfo{person}{Chunyuan Li}, \bibinfo{person}{Qingyang Wu}, {and} \bibinfo{person}{Yong~Jae Lee}.} \bibinfo{year}{2023}\natexlab{b}.
\newblock \bibinfo{title}{Visual Instruction Tuning}.
\newblock
\newblock
\showeprint[arxiv]{2304.08485}


\bibitem[Liu et~al\mbox{.}(2022)]%
        {liu2022convnet}
\bibfield{author}{\bibinfo{person}{Zhuang Liu}, \bibinfo{person}{Hanzi Mao}, \bibinfo{person}{Chao-Yuan Wu}, \bibinfo{person}{Christoph Feichtenhofer}, \bibinfo{person}{Trevor Darrell}, {and} \bibinfo{person}{Saining Xie}.} \bibinfo{year}{2022}\natexlab{}.
\newblock \showarticletitle{A convnet for the 2020s}. In \bibinfo{booktitle}{\emph{Proceedings of the IEEE/CVF conference on computer vision and pattern recognition}}. \bibinfo{pages}{11976--11986}.
\newblock


\bibitem[Lloyd(1982)]%
        {lloyd1982least}
\bibfield{author}{\bibinfo{person}{Stuart Lloyd}.} \bibinfo{year}{1982}\natexlab{}.
\newblock \showarticletitle{Least squares quantization in PCM}.
\newblock \bibinfo{journal}{\emph{IEEE transactions on information theory}} \bibinfo{volume}{28}, \bibinfo{number}{2} (\bibinfo{year}{1982}), \bibinfo{pages}{129--137}.
\newblock


\bibitem[Loshchilov and Hutter(2017)]%
        {loshchilov2017decoupled}
\bibfield{author}{\bibinfo{person}{Ilya Loshchilov} {and} \bibinfo{person}{Frank Hutter}.} \bibinfo{year}{2017}\natexlab{}.
\newblock \bibinfo{title}{Decoupled weight decay regularization}.
\newblock
\newblock
\showeprint[arxiv]{1711.05101}


\bibitem[Natarajan et~al\mbox{.}(2013)]%
        {natarajan2013learning}
\bibfield{author}{\bibinfo{person}{Nagarajan Natarajan}, \bibinfo{person}{Inderjit~S Dhillon}, \bibinfo{person}{Pradeep~K Ravikumar}, {and} \bibinfo{person}{Ambuj Tewari}.} \bibinfo{year}{2013}\natexlab{}.
\newblock \showarticletitle{Learning with noisy labels}. In \bibinfo{booktitle}{\emph{Advances in neural information processing systems}}, Vol.~\bibinfo{volume}{26}.
\newblock


\bibitem[Ogawa et~al\mbox{.}(2018)]%
        {ogawa2018object}
\bibfield{author}{\bibinfo{person}{Toru Ogawa}, \bibinfo{person}{Atsushi Otsubo}, \bibinfo{person}{Rei Narita}, \bibinfo{person}{Yusuke Matsui}, \bibinfo{person}{Toshihiko Yamasaki}, {and} \bibinfo{person}{Kiyoharu Aizawa}.} \bibinfo{year}{2018}\natexlab{}.
\newblock \bibinfo{title}{Object detection for comics using manga109 annotations}.
\newblock
\newblock
\showeprint[arxiv]{1803.08670}


\bibitem[OpenAI(2022)]%
        {chatgpt}
\bibfield{author}{\bibinfo{person}{OpenAI}.} \bibinfo{year}{2022}\natexlab{}.
\newblock \bibinfo{title}{ChatGPT}.
\newblock \bibinfo{howpublished}{\url{https://openai.com/blog/chatgpt/}}.
\newblock


\bibitem[OpenAI(2023)]%
        {gpt4}
\bibfield{author}{\bibinfo{person}{OpenAI}.} \bibinfo{year}{2023}\natexlab{}.
\newblock \bibinfo{title}{GPT-4 Technical Report}.
\newblock
\newblock
\showeprint[arxiv]{2303.08774}


\bibitem[Ren et~al\mbox{.}(2015)]%
        {ren2015faster}
\bibfield{author}{\bibinfo{person}{Shaoqing Ren}, \bibinfo{person}{Kaiming He}, \bibinfo{person}{Ross Girshick}, {and} \bibinfo{person}{Jian Sun}.} \bibinfo{year}{2015}\natexlab{}.
\newblock \showarticletitle{Faster r-cnn: Towards real-time object detection with region proposal networks}.
\newblock \bibinfo{journal}{\emph{Advances in neural information processing systems}}  \bibinfo{volume}{28} (\bibinfo{year}{2015}).
\newblock


\bibitem[Rigaud et~al\mbox{.}(2015)]%
        {rigaud2015speech}
\bibfield{author}{\bibinfo{person}{Christophe Rigaud}, \bibinfo{person}{Nam Le~Thanh}, \bibinfo{person}{J-C Burie}, \bibinfo{person}{J-M Ogier}, \bibinfo{person}{Motoi Iwata}, \bibinfo{person}{Eiki Imazu}, {and} \bibinfo{person}{Koichi Kise}.} \bibinfo{year}{2015}\natexlab{}.
\newblock \showarticletitle{Speech balloon and speaker association for comics and manga understanding}. In \bibinfo{booktitle}{\emph{Proceedings of 13th international conference on document analysis and recognition}}. IEEE, \bibinfo{pages}{351--355}.
\newblock


\bibitem[Song et~al\mbox{.}(2022)]%
        {song2022learning}
\bibfield{author}{\bibinfo{person}{Hwanjun Song}, \bibinfo{person}{Minseok Kim}, \bibinfo{person}{Dongmin Park}, \bibinfo{person}{Yooju Shin}, {and} \bibinfo{person}{Jae-Gil Lee}.} \bibinfo{year}{2022}\natexlab{}.
\newblock \showarticletitle{Learning from noisy labels with deep neural networks: A survey}.
\newblock \bibinfo{journal}{\emph{IEEE transactions on neural networks and learning systems}} (\bibinfo{year}{2022}).
\newblock


\bibitem[Tang et~al\mbox{.}(2020)]%
        {tang2020unbiased}
\bibfield{author}{\bibinfo{person}{Kaihua Tang}, \bibinfo{person}{Yulei Niu}, \bibinfo{person}{Jianqiang Huang}, \bibinfo{person}{Jiaxin Shi}, {and} \bibinfo{person}{Hanwang Zhang}.} \bibinfo{year}{2020}\natexlab{}.
\newblock \showarticletitle{Unbiased scene graph generation from biased training}. In \bibinfo{booktitle}{\emph{Proceedings of the IEEE conference on computer vision and pattern recognition}}. \bibinfo{pages}{3716--3725}.
\newblock


\bibitem[Tsubota et~al\mbox{.}(2018)]%
        {tsubota2018adaptation}
\bibfield{author}{\bibinfo{person}{Koki Tsubota}, \bibinfo{person}{Toru Ogawa}, \bibinfo{person}{Toshihiko Yamasaki}, {and} \bibinfo{person}{Kiyoharu Aizawa}.} \bibinfo{year}{2018}\natexlab{}.
\newblock \showarticletitle{Adaptation of Manga Face Representation for Accurate Clustering}. In \bibinfo{booktitle}{\emph{SIGGRAPH Asia 2018 Posters}}. \bibinfo{publisher}{Association for Computing Machinery}, Article \bibinfo{articleno}{15}, \bibinfo{numpages}{2}~pages.
\newblock


\bibitem[Wei et~al\mbox{.}(2022)]%
        {wei2022emergent}
\bibfield{author}{\bibinfo{person}{Jason Wei}, \bibinfo{person}{Yi Tay}, \bibinfo{person}{Rishi Bommasani}, \bibinfo{person}{Colin Raffel}, \bibinfo{person}{Barret Zoph}, \bibinfo{person}{Sebastian Borgeaud}, \bibinfo{person}{Dani Yogatama}, \bibinfo{person}{Maarten Bosma}, \bibinfo{person}{Denny Zhou}, \bibinfo{person}{Donald Metzler}, {et~al\mbox{.}}} \bibinfo{year}{2022}\natexlab{}.
\newblock \showarticletitle{Emergent abilities of large language models}.
\newblock \bibinfo{journal}{\emph{Transactions on Machine Learning Research}} (\bibinfo{year}{2022}).
\newblock


\bibitem[Xu et~al\mbox{.}(2017)]%
        {xu2017scene}
\bibfield{author}{\bibinfo{person}{Danfei Xu}, \bibinfo{person}{Yuke Zhu}, \bibinfo{person}{Christopher~B Choy}, {and} \bibinfo{person}{Li Fei-Fei}.} \bibinfo{year}{2017}\natexlab{}.
\newblock \showarticletitle{Scene graph generation by iterative message passing}. In \bibinfo{booktitle}{\emph{Proceedings of the IEEE conference on computer vision and pattern recognition}}. \bibinfo{pages}{5410--5419}.
\newblock


\bibitem[Zellers et~al\mbox{.}(2018)]%
        {zellers2018neural}
\bibfield{author}{\bibinfo{person}{Rowan Zellers}, \bibinfo{person}{Mark Yatskar}, \bibinfo{person}{Sam Thomson}, {and} \bibinfo{person}{Yejin Choi}.} \bibinfo{year}{2018}\natexlab{}.
\newblock \showarticletitle{Neural motifs: Scene graph parsing with global context}. In \bibinfo{booktitle}{\emph{Proceedings of the IEEE conference on computer vision and pattern recognition}}. \bibinfo{pages}{5831--5840}.
\newblock


\bibitem[Zhang et~al\mbox{.}(2023)]%
        {zhang2023benchmarking}
\bibfield{author}{\bibinfo{person}{Tianyi Zhang}, \bibinfo{person}{Faisal Ladhak}, \bibinfo{person}{Esin Durmus}, \bibinfo{person}{Percy Liang}, \bibinfo{person}{Kathleen McKeown}, {and} \bibinfo{person}{Tatsunori~B. Hashimoto}.} \bibinfo{year}{2023}\natexlab{}.
\newblock \bibinfo{title}{Benchmarking Large Language Models for News Summarization}.
\newblock
\newblock
\showeprint[arxiv]{2301.13848}


\bibitem[Zhang et~al\mbox{.}(2022)]%
        {zhang2022unsupervised}
\bibfield{author}{\bibinfo{person}{Zhimin Zhang}, \bibinfo{person}{Zheng Wang}, {and} \bibinfo{person}{Wei Hu}.} \bibinfo{year}{2022}\natexlab{}.
\newblock \bibinfo{title}{Unsupervised Manga Character Re-identification via Face-body and Spatial-temporal Associated Clustering}.
\newblock
\newblock
\showeprint[arxiv]{2204.04621}


\bibitem[Zheng et~al\mbox{.}(2020)]%
        {zheng2020cartoon}
\bibfield{author}{\bibinfo{person}{Yi Zheng}, \bibinfo{person}{Yifan Zhao}, \bibinfo{person}{Mengyuan Ren}, \bibinfo{person}{He Yan}, \bibinfo{person}{Xiangju Lu}, \bibinfo{person}{Junhui Liu}, {and} \bibinfo{person}{Jia Li}.} \bibinfo{year}{2020}\natexlab{}.
\newblock \showarticletitle{Cartoon face recognition: A benchmark dataset}. In \bibinfo{booktitle}{\emph{Proceedings of the 28th ACM international conference on multimedia}}. \bibinfo{pages}{2264--2272}.
\newblock


\bibitem[Zhu et~al\mbox{.}(2023)]%
        {zhu2023minigpt}
\bibfield{author}{\bibinfo{person}{Deyao Zhu}, \bibinfo{person}{Jun Chen}, \bibinfo{person}{Xiaoqian Shen}, \bibinfo{person}{Xiang Li}, {and} \bibinfo{person}{Mohamed Elhoseiny}.} \bibinfo{year}{2023}\natexlab{}.
\newblock \bibinfo{title}{Minigpt-4: Enhancing vision-language understanding with advanced large language models}.
\newblock
\newblock
\showeprint[arxiv]{2304.10592}


\end{thebibliography}

\appendix
\clearpage
\newcommand\beginsupplement{%
        \setcounter{table}{0}
        \renewcommand{\thetable}{\Alph{table}}%
        \setcounter{figure}{0}
        \renewcommand{\thefigure}{\Alph{figure}}%
     }
\beginsupplement

%%%%%%%%%%%%%%%%%%%%%%%%%%%%%%%%%%%%%%%%%%%%%%%%%%%%%%%%%%%%%%%%
\section{LLM Prompts}
%%%%%%%%%%%%%%%%%%%%%%%%%%%%%%%%%%%%%%%%%%%%%%%%%%%%%%%%%%%%%%%%
In this section, we show the prompts of large language models (LLMs) for speaker prediction. 
Through our experiments, we observed that using prompts in Japanese leads to higher prediction accuracy than using prompts in English. This is probably because the input texts are in Japanese; using prompts in the same language could reduce confusion for the LLMs.
For the understanding of our prompts, we have translated them into English.
The introduced prompts are fed into GPT-4 as system prompts while user prompts only contain the texts in each comic.

\subsection{Character Name Extraction}
The prompt we used for extracting character names from dialogues is shown below.
Since the names that appear in dialogues might be incomplete, potentially leading to multiple names being extracted for the same character, we instructed the LLMs to use contextual information to infer and output full names wherever possible.

\begin{lstlisting}
Given a sequence of manga text in Japanese, identify the names of the characters estimated to appear.

### Note
- When extracting character names, provide full names if possible, e.g., "Taro Yamada".
- If full names are not explicitly mentioned, analyze the context within the text to deduce the full names.
- If the name of a character is unknown, describe them by their occupation or their relationship with other characters, e.g., "the teacher" or "Yamada's mother".

### Input/Output format
[Input format]
Text ID | Text

[Output format]
Character ID | Character Name
\end{lstlisting}

\subsection{Context Extraction}
The prompt we used for extracting a story summary and character profiles is shown below.
Here, we take the prompt for \texttt{LoveHina vol01} as an example.
We instructed the LLMs to output the context information in Japanese to make the language consistent as mentioned above.

\begin{lstlisting}
Given a sequence of manga text in Japanese and a list of characters who appear in it, produce a story summary and character profiles based on the following steps. 
Note: The output should be in Japanese.

1. Summary: summarize the story in the manga.
2. Characters: For each character listed, provide details about their attributes, including gender, estimated age, role, a brief description, and relationships with other characters.

### List of appearing characters
Character ID | Character Name
A | Keitaro
B | Naru
...
    
### Input/Output format
[Input format]
Text ID | Text

[Output format]
1. Summary:
2. Characters:
- Keitaro:
- Naru: 
- ...
\end{lstlisting}

\subsection{Initial Speaker Prediction (w/o candidates)}
The prompt of the initial speaker prediction based only on the texts is shown below.
We instructed the LLMs to output not only the predicted speaker's name but also a confidence level for that prediction.
The data with a low confidence level is not used in the subsequent steps.
We defined five confidence levels and provided detailed explanations for the criteria.

\begin{lstlisting}
Given a sequence of manga text in Japanese and a list of characters who appear in it, predict the speaker of each text considering the context information. 
Please also output a confidence level of prediction on a scale of 5.

### Note
- Not all the given characters might be speaking in the provided text.
- The number of output lines should be the same as the number of input lines.

### List of appearing characters
Character ID | Character Name
A | Keitaro
B | Naru
...

### Context information
1. Summary: Keitaro tries to ...
2. Characters:
- Keitaro: Main character who ...
- Naru: Heroine of the story. ...
- ...

### Input/Output format
[Input format]
Text ID | Text

[Output format]
Text ID | Character Name | Character ID | Confidence Level

### Confidence level
Score and criteria:
1: Completely uncertain, the prediction is near random.
2: Low confidence, the probability that the prediction is correct is less than 50%.
3: Moderate confidence, the prediction is likely correct but could be wrong.
4: High confidence, the prediction is probably correct, but not 100% certain.
5: Very high confidence, the prediction is almost certainly correct.

### Input/Output example
[Input]
1 | Hey, Naru.
2 | What, Keitaro?

[Output]
1 | Keitaro | A | 5
2 | Naru | B | 4
\end{lstlisting}

\subsection{Iterative Speaker Prediction (w/ candidates)}
The prompt of the iterative speaker prediction using the speaker candidates is shown below.
A speaker candidate is obtained for each text based on the character identification results.
We integrated image-based predictions and textual information by providing LLMs with speaker candidates.
We also supplied the prediction probability for each speaker candidate so that LLMs can use the information, which leads to 2.0\% improvement in accuracy as shown in Table 3 of the main paper.
% Additionally, in the prompt, we stated that the speaker candidates might not always be correct.

\begin{lstlisting}
Given a sequence of manga text in Japanese and a list of characters who appear in it, predict the speaker of each text considering the context information.
Please also output a confidence level of prediction on a scale of 5.

For each text, you will be given a speaker candidate, which is obtained from the image-based prediction, along with a probability for that prediction. Use this as a reference.

### Note
- Not all the given characters might be speaking in the provided text.
- The number of output lines should be the same as the number of input lines.
- If no image-based predictions are given, predict the speaker based on the text and the context.
- The image-based predictions may not always be correct. Exercise caution, especially when the prediction probability is low.


### List of appearing characters
Character ID | Character Name
A | Keitaro
B | Naru
...

### Context information
1. Summary: Keitaro tries to ...
2. Characters:
- Keitaro: Main character who ...
- Naru: Heroine of the story. ...
- ...

### Input/Output format
[Input format]
Text ID | Text | Speaker Candidate (Prediction Probability)

[Output format]
Text ID | Character Name | Character ID | Confidence Level

### Confidence level
Score and criteria:
1: Completely uncertain, the prediction is near random.
2: Low confidence, the probability that the prediction is correct is less than 50%.
3: Moderate confidence, the prediction is likely correct but could be wrong.
4: High confidence, the prediction is probably correct, but not 100% certain.
5: Very high confidence, the prediction is almost certainly correct.

### Input/Output example
[Input]
1 | Hey, Naru. | Keitaro (0.56)
2 | What, Keitaro? | Naru (0.8)

[Output]
1 | Keitaro | A | 5
2 | Naru | B | 4
\end{lstlisting}

%%%%%%%%%%%%%%%%%%%%%%%%%%%%%%%%%%%%%%%%%%%%%%%%%%%%%%%%%%%%%%%%
\section{Character Region Classifier}
%%%%%%%%%%%%%%%%%%%%%%%%%%%%%%%%%%%%%%%%%%%%%%%%%%%%%%%%%%%%%%%%
This section describes the details of character region classification. We utilize the ResNet50 model~\cite{he2016deep}, adapted to classify character regions. Our training method consists of a pre-training phase on general comics and a fine-tuning phase for individual unseen comics in the test set. The steps of pre-training, fine-tuning, and testing are detailed below.

\noindentparagraph{\textbf{Pre-training.}}
We pre-train the models for comic character classification using 
Manga109 dataset~\cite{ogawa2018object,aizawa2020building}.
% Because of the limited data for each individual comic in fine-tuning step, 
% learning good representation by pre-training is critical.
The model are initialized with weights from a model trained on ImageNet~\cite{deng2009imagenet,he2016deep}.
The training set includes 349 characters from the Manga109 training set, which consists of separated titles from the test set.
Character regions are cropped and resized to 270$\times$270, followed by data augmentation including random 256$\times$256 crops, horizontal flips, and rotations. Training lasts for 50 epochs using AdamW optimizer~\cite{loshchilov2017decoupled} with a batch size of 32.
The learning rate is set to $1 \times 10^{-4}$, and it is reduced by a factor of 0.1 after the 20th and 40th epochs.

\noindentparagraph{\textbf{Fine-tuning.}}
In the character identification step of our framework, the pre-trained model is fine-tuned for each comic in the test set using the pseudo labels generated from speaker prediction results.
We split the whole data into train and validation sets, allocating 10\% of the data randomly for validation.
We run fine-tuning with 10 epochs and select the model with the highest validation accuracy.
The learning rate is set to $1 \times 10^{-4}$, with other parameters and augmentation strategies being consistent with the pre-training phase.

\noindentparagraph{\textbf{Testing.}}
In testing, we used 10-crop testing~\cite{krizhevsky2012imagenet} with crops sized at 256$\times$256 from the image resized to 270$\times$270. To handle the instability of training with noisy labels, we used a simple ensemble approach: we trained five models using the same training data, where the output probability of classification is calculated by averaging the outputs from all models.
As future work, we might be able to employ specific methods to handle noisy labels to improve the results~\cite{natarajan2013learning,song2022learning}.

%%%%%%%%%%%%%%%%%%%%%%%%%%%%%%%%%%%%%%%%%%%%%%%%%%%%%%%%%%%%%%%%
\section{Main Results}
%%%%%%%%%%%%%%%%%%%%%%%%%%%%%%%%%%%%%%%%%%%%%%%%%%%%%%%%%%%%%%%%
In this section, we present more examples of our experimental results, which were obtained using the same experimental settings as the results of Table 1 in the main paper. 
As mentioned in Section 4.2, to validate the effectiveness of our proposed iterative multimodal fusion method, we conducted evaluations in two aspects: \textbf{Unimodal vs. Multimodal} and \textbf{One-Step vs. Iterative}.
Additionally, we analyze the limitations of our current method through several failure examples to show the direction of future work.

\subsection{Unimodal vs. Multimodal}
\begin{figure*}[t]
    \centering
    \includegraphics[width=\linewidth]{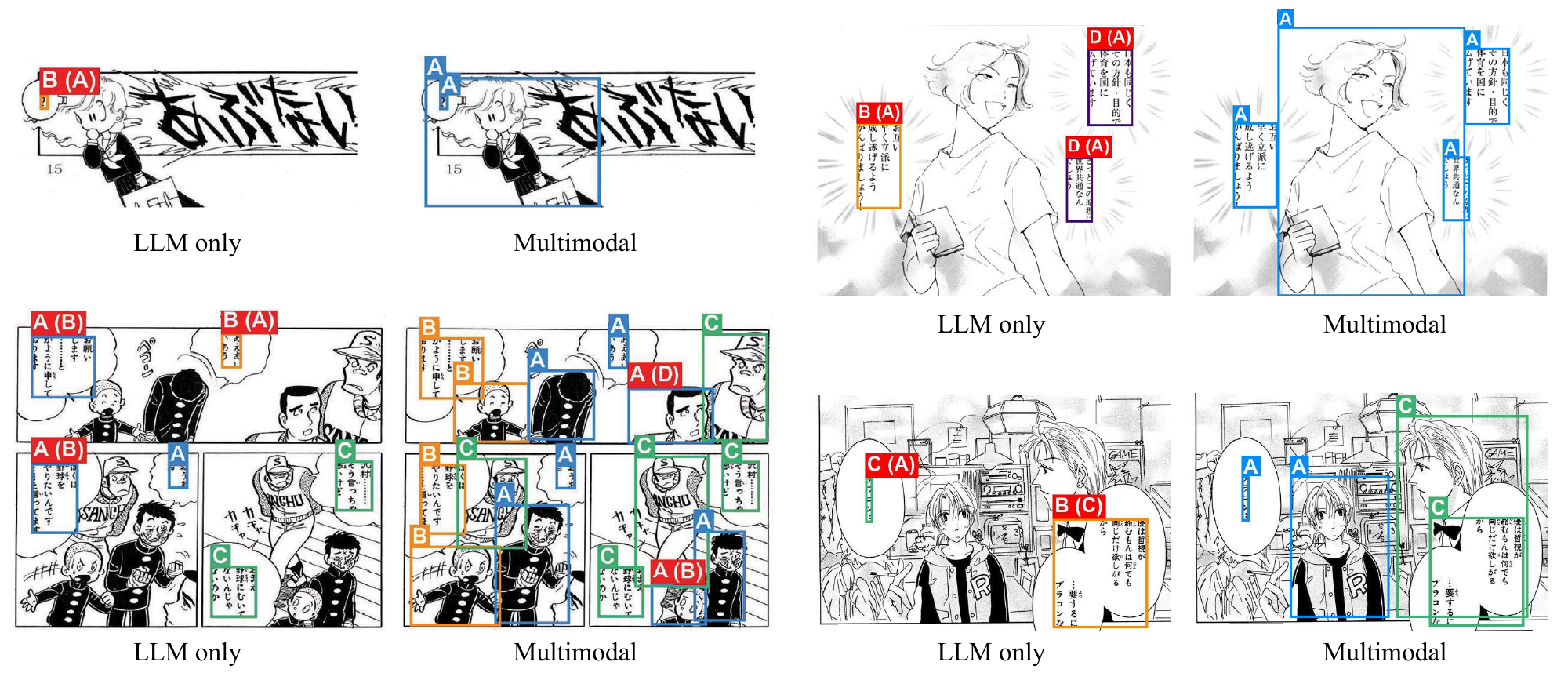}
    \caption{Speaker prediction results obtained using a single modality (textual information) and multiple modalities combined. 
    Color of the bounding box indicates the predicted character label. Red labels indicate failure predictions.
    Courtesy of Tashiro Kimu, Hikochi Sakuya, Yoshimori Mikio, Karikawa Seyu.}
    \label{fig:case1}
\end{figure*}

\begin{figure*}[t]
    \centering
    \includegraphics[width=\linewidth]{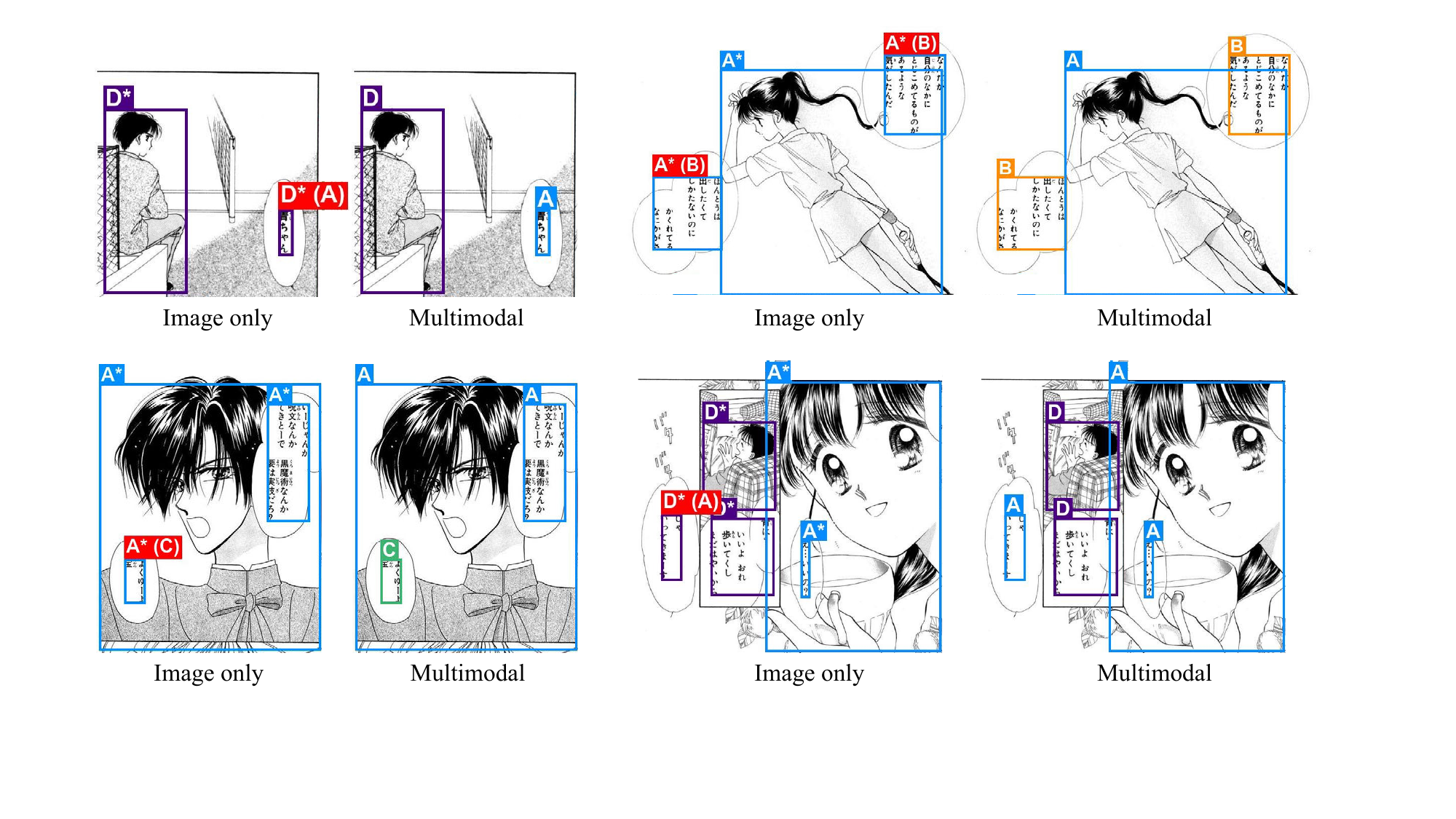}
    \caption{Speaker prediction results obtained using a single modality (visual information) and multiple modalities combined.
    The labels on the boxes (e.g., `A') are character labels. 
    Labels in brackets are the ground truth. (e.g., `A (B)' is the case where the ground truth is B but the prediction is A.)
    $^*$ indicates that the image-based method used the ground truth of character labels.
    Courtesy of Ayumi Yui, Hanada Sakumi.
    }
    \label{fig:case2}
\end{figure*}

To show the effectiveness of using multimodal information,
we compare our proposed multimodal method with the methods using only visual information or textual content, as shown in Figure~\ref{fig:case1} and Figure~\ref{fig:case2}. 
Color of the bounding box indicates the predicted character label. Red labels indicate failure predictions.
The results of \textit{LLM only} and \textit{K-means+SGG} baseline are shown as text-only and image-only results, respectively.
As explained in the main paper, \textit{K-means+SGG} used the ground truth to map each cluster into character labels because the image-only approach cannot identify character labels.
For the \textit{LLM only} method, we did not perform character identification.

From Figure~\ref{fig:case1}, we can observe that LLMs struggle to make accurate predictions in the case that the texts lack distinctive character features. However, when combining these with image-based predictions, visual information provides essential cues about the speakers, enabling the model to make correct predictions.
The results in  Figure~\ref{fig:case2} show that an image-only method failed to predict the case that the speaker is not the character closest to the text. In contrast, the multimodal method can predict correctly by using textual content such as story context.
These results show the effectiveness of using multimodal information for these tasks,
which accords with how humans do when reading comics, i.e.,
identifying the speaker of dialogue using both visual and textual information.

\subsection{One-Step vs. Iterative}
\begin{figure*}[h]
    \centering
    \begin{minipage}[c]{\linewidth}
        \centering
        \includegraphics[width=0.99\linewidth]{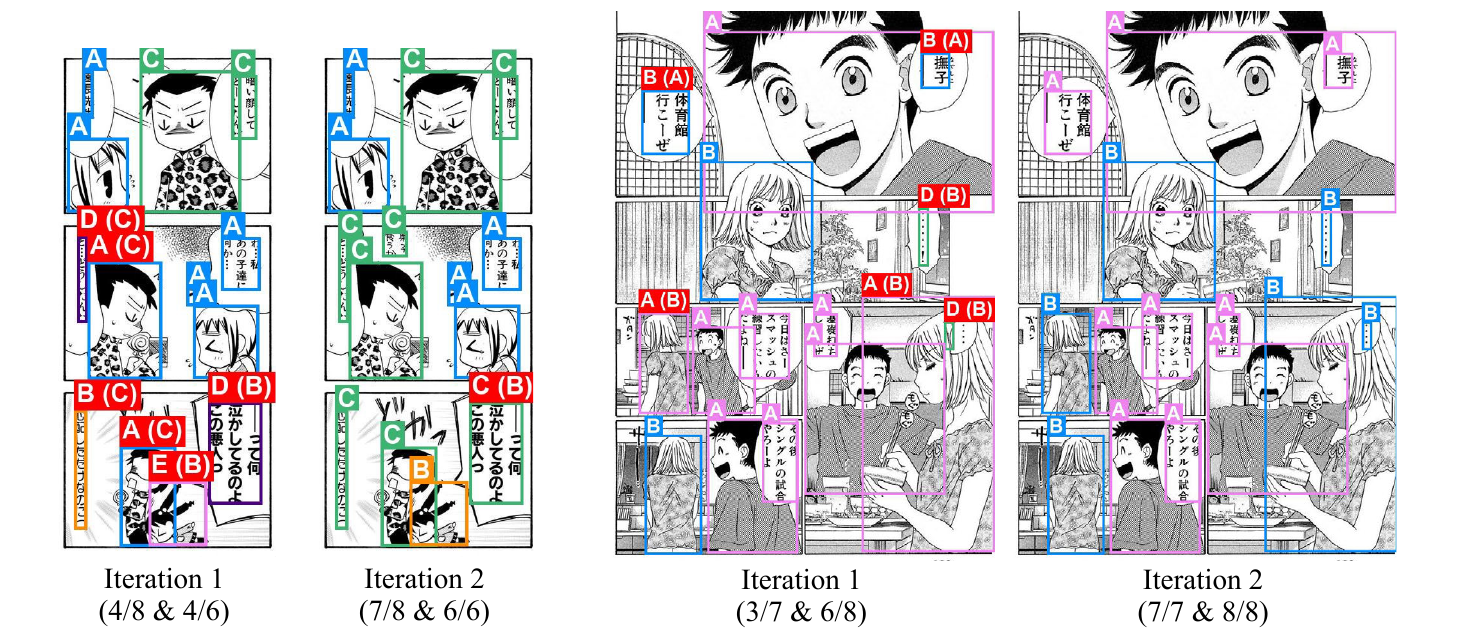}
    \end{minipage}
    \begin{minipage}[c]{\linewidth}
        \centering
        \includegraphics[width=0.99\linewidth]{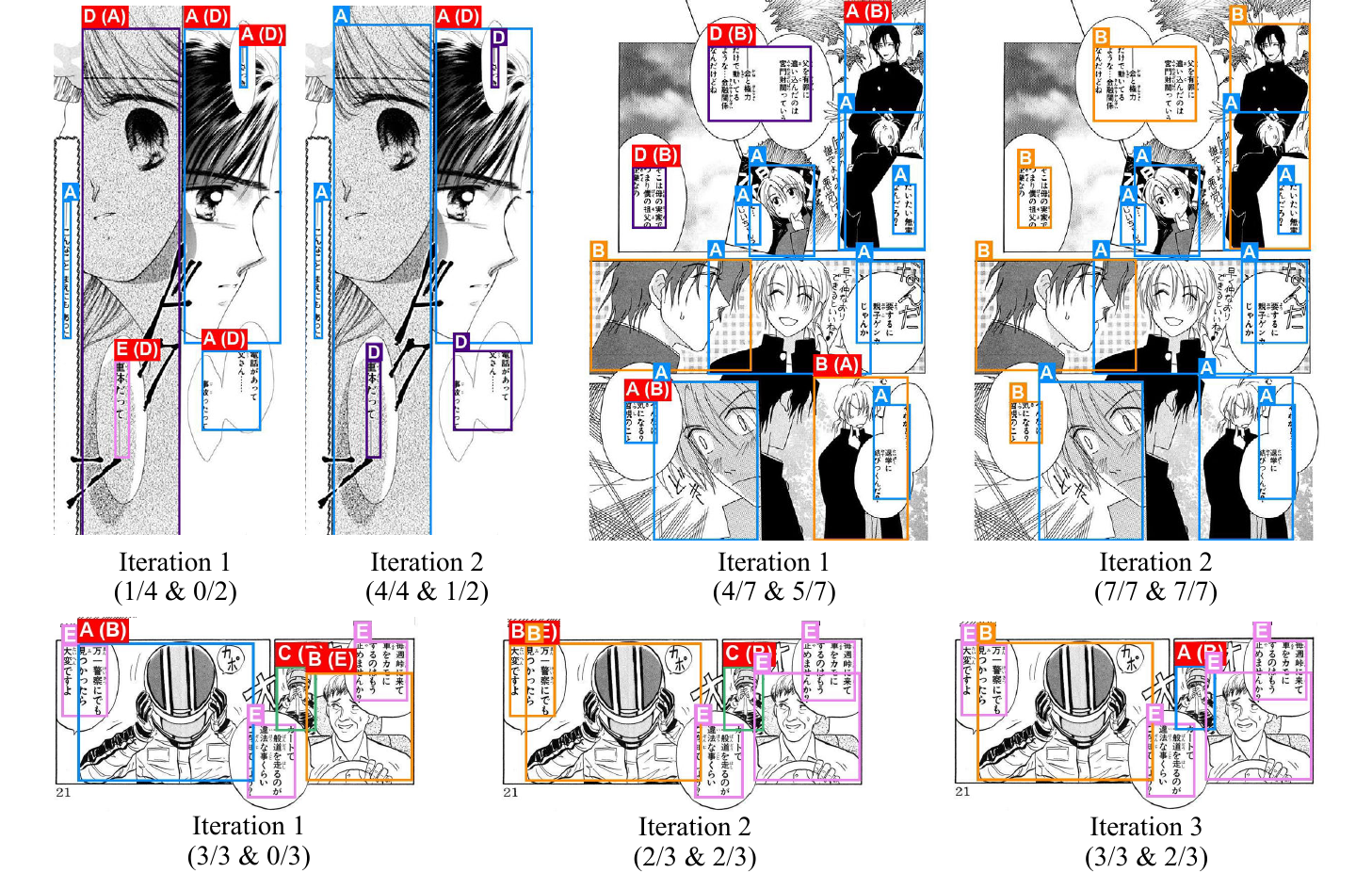}
    \end{minipage}
    \caption{Results of character identification and speaker prediction across different iterations. (Accuracy of speaker pred. \& character id.). Courtesy of Tenya, Saki Kaori, Ayumi Yui, Karikawa Seyu, Matsuda Naomasa.}
    \label{fig:case3_part1}
\end{figure*}

\begin{figure*}[h]
    \centering
    \begin{minipage}[c]{\linewidth}
        \centering
        \includegraphics[width=\linewidth]{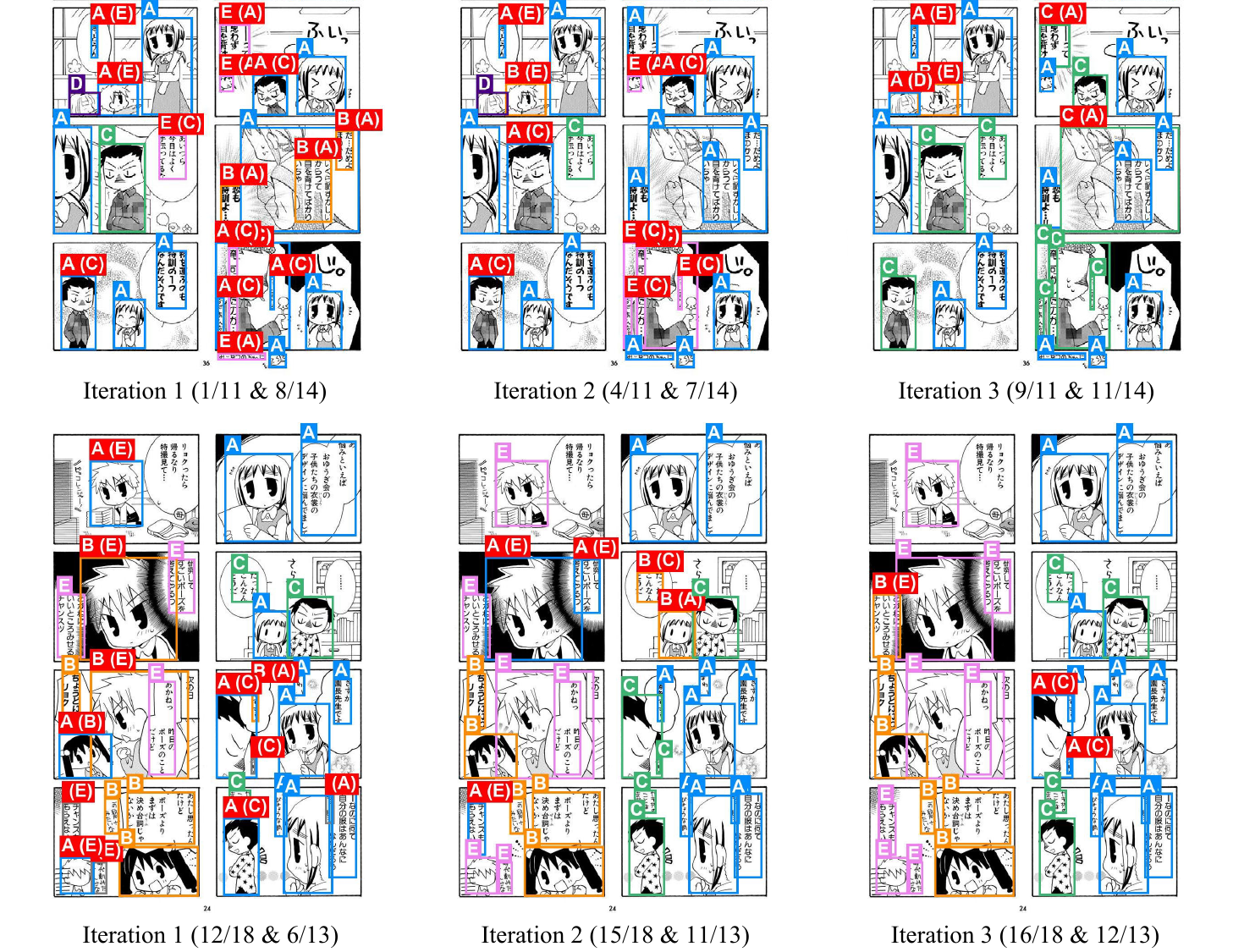}
    \end{minipage}
    \begin{minipage}[c]{\linewidth}
        \centering
        \includegraphics[width=\linewidth]{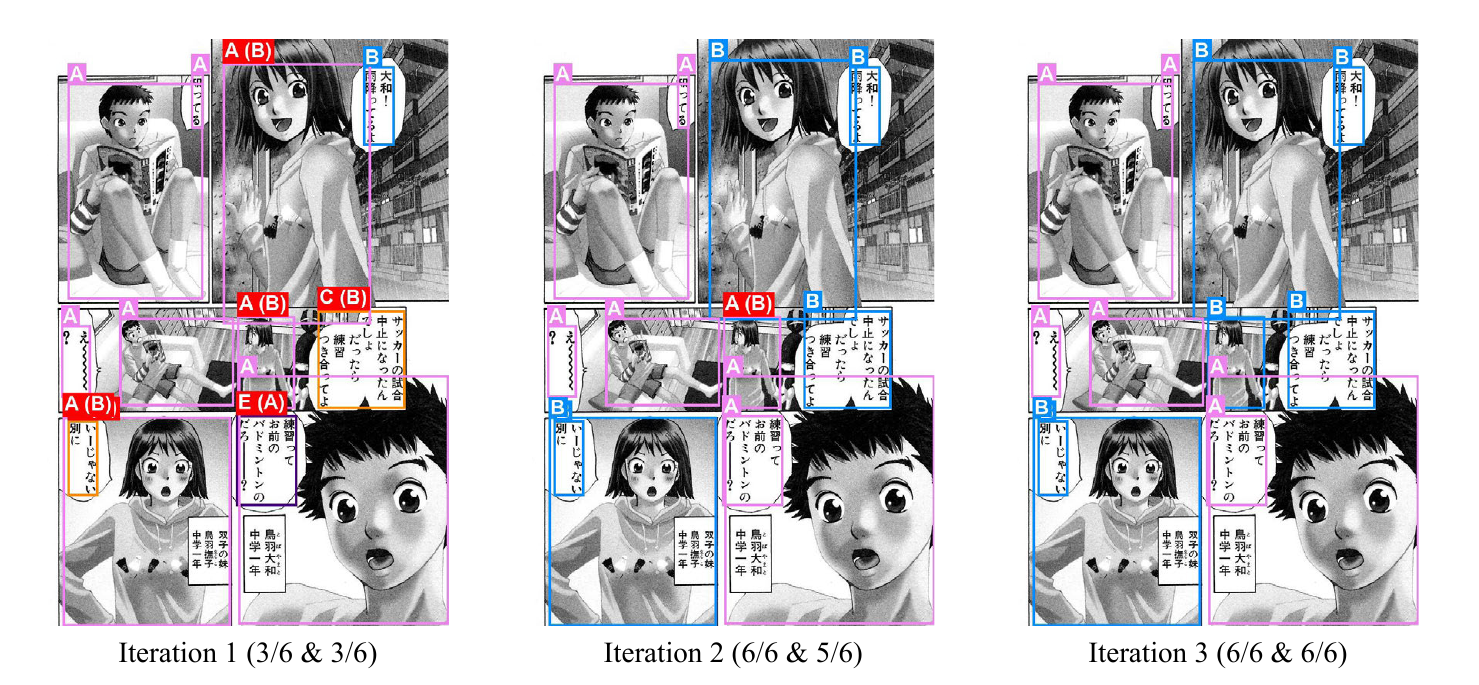}
    \end{minipage}
    \caption{Results of character identification and speaker prediction across different iterations. (Accuracy of speaker pred. \& character id.). Courtesy of Tenya, Saki Kaori.}
    \label{fig:case3_part2}
\end{figure*}

Experimental results of character identification and speaker prediction under different iteration times are shown in Figure~\ref{fig:case3_part1} and Figure~\ref{fig:case3_part2}.
The accuracy for both character identification and speaker prediction improves as the number of iterations increases.
This demonstrates that enhanced results of character identification can positively influence the accuracy of speaker prediction, and vice versa, further validating the effectiveness of our iterative approach.

\clearpage

\clearpage

\clearpage
\subsection{Failure Examples}
\begin{figure}[h]
    \centering
    \includegraphics[width=\linewidth]{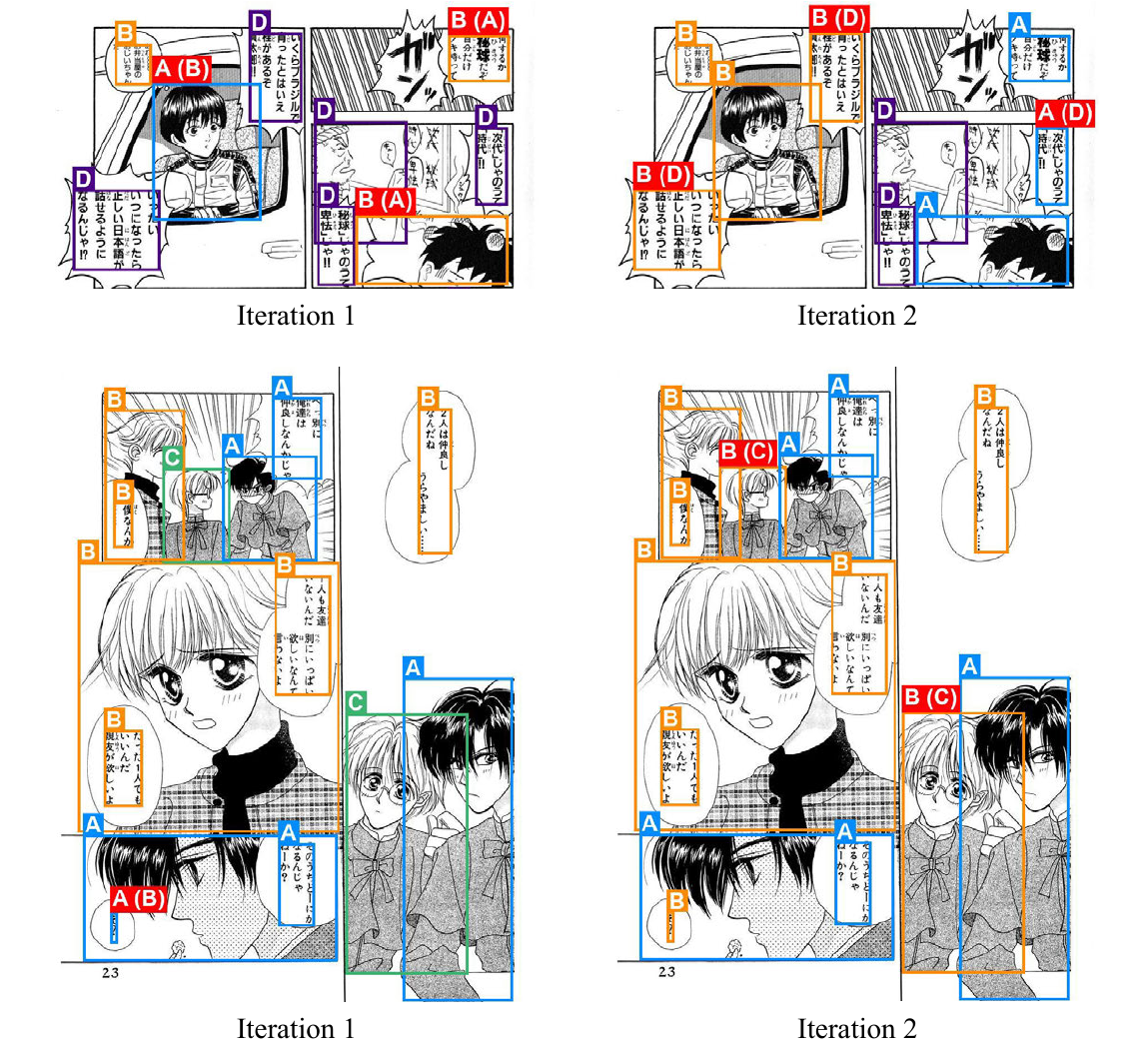}
    \caption{Failure examples where prediction results deteriorate with increasing iterations. Courtesy of Matsuda Naomasa, Hanada Sakumi.}
    \label{fig:failure_iter}
\end{figure}

Figure~\ref{fig:failure_iter} shows the cases where increasing the number of iterations leads to poorer prediction results.
In the first example, for the texts positioned on the left (purple box with label \texttt{D}), the LLMs initially made correct speaker prediction in iteration 1. 
However, since character \texttt{B} is closer to these texts, its label is propagated to these texts, leading to incorrect predictions in iteration 2. 
% This example indicates that the results of character identification can potentially have a negative impact on speaker prediction.
In the second example, the character classifier originally made the correct identification. 
However,
labels of the character \texttt{C} are changed to incorrect label \texttt{B}. This is because the labels to dialogue are propagated to characters even if the character does not speak in this figure.
% This indicates that the results of speaker prediction might have a negative impact on character identification.

These examples show that improving the accuracy of either character identification or speaker prediction does not necessarily enhance the performance of the other in the case that the correspondence between text and character regions is not clear.
It suggests future directions such as a method to generate reliable pseudo-labels 
or a robust training method to handle noisy labels.

\section{Zero-shot Results}
\begin{figure}[H]
    \centering
    \begin{minipage}[c]{\linewidth}
        \centering
        \fbox{\includegraphics[width=0.85\linewidth]{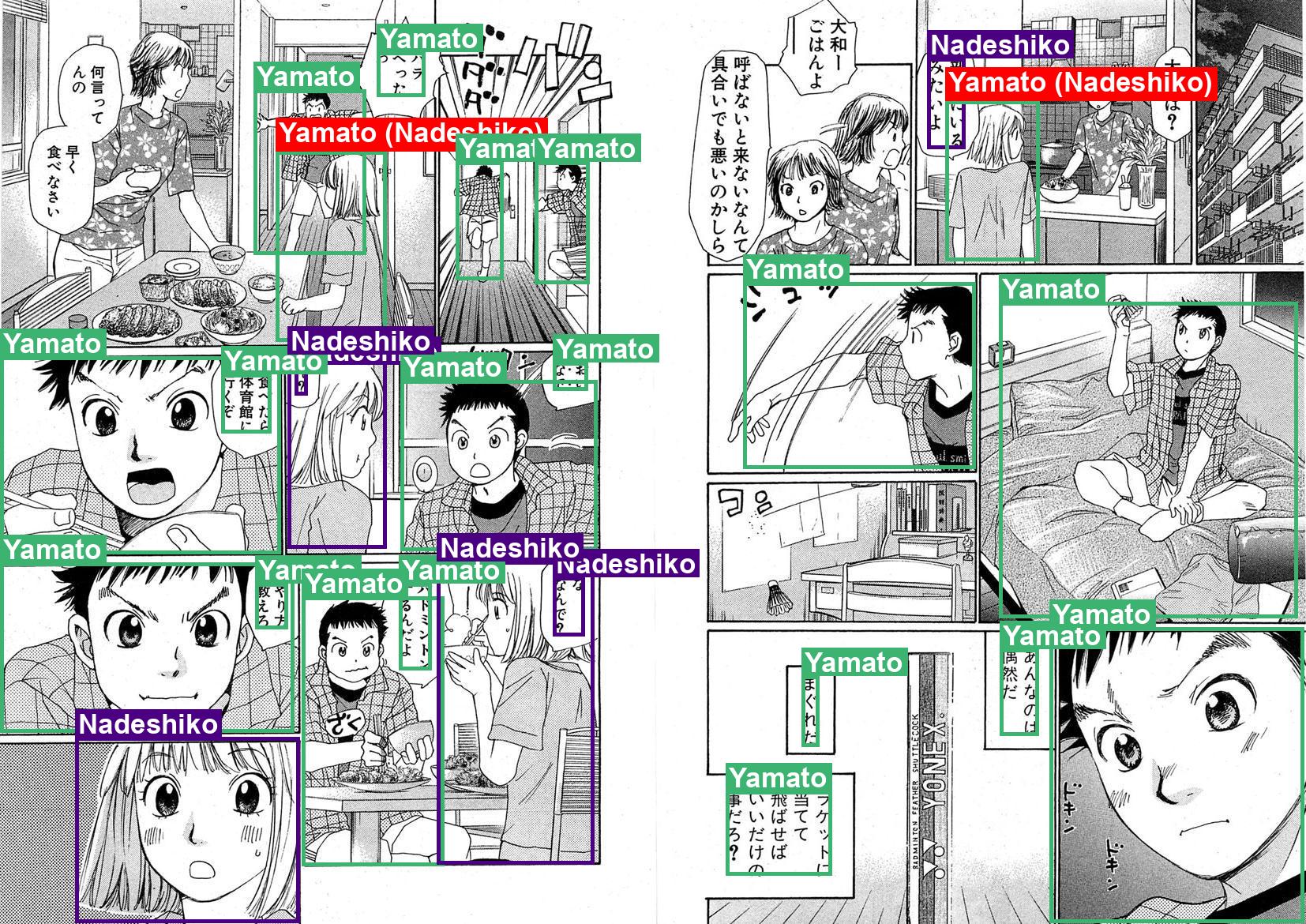}}
    \end{minipage}%
    \vspace{15pt}
    \begin{minipage}[c]{\linewidth}
        \centering
        \fbox{\includegraphics[width=0.85\linewidth]{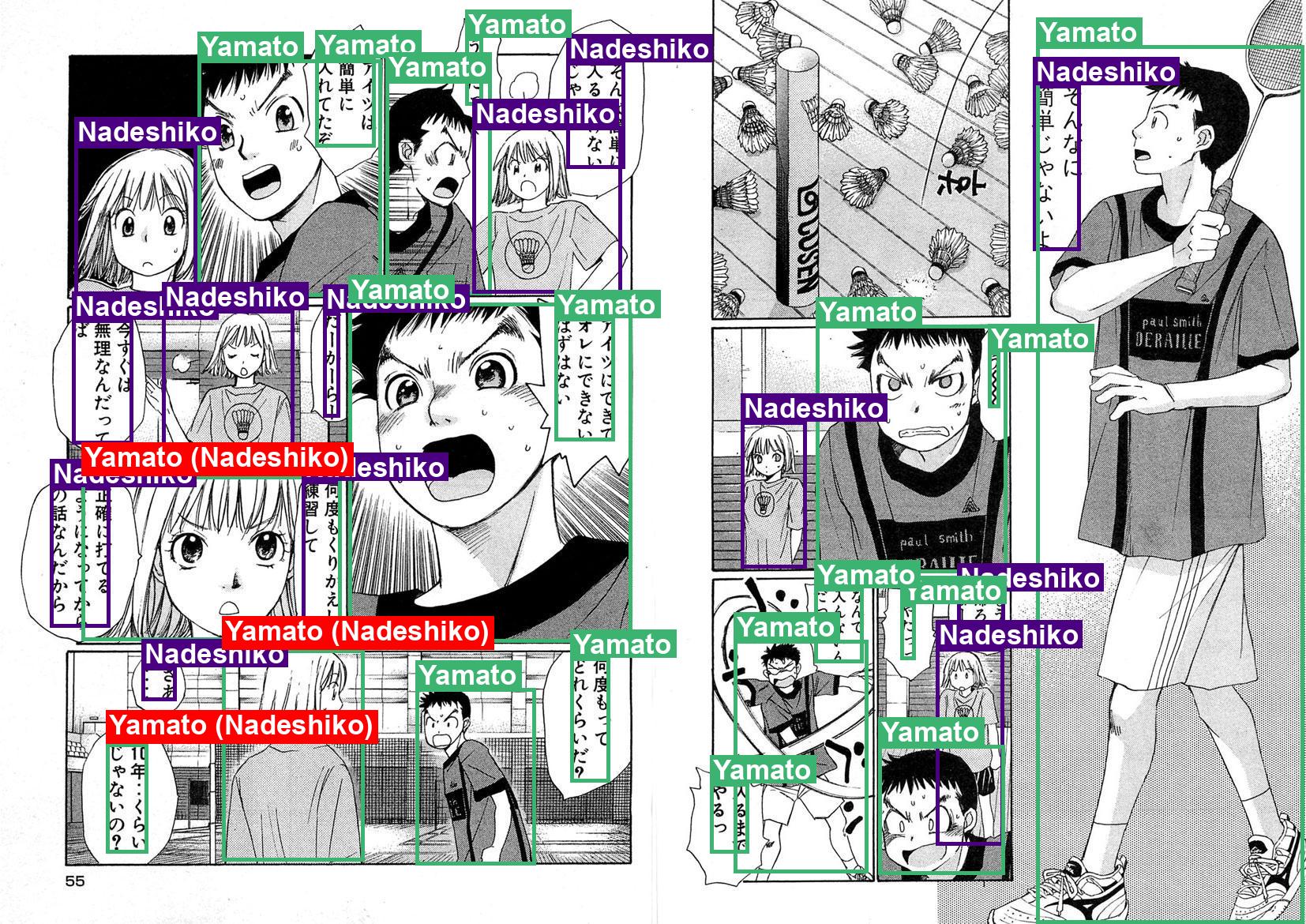}}
    \end{minipage}%
    \vspace{15pt}
    \begin{minipage}[c]{\linewidth}
        \centering
        \fbox{\includegraphics[width=0.85\linewidth]{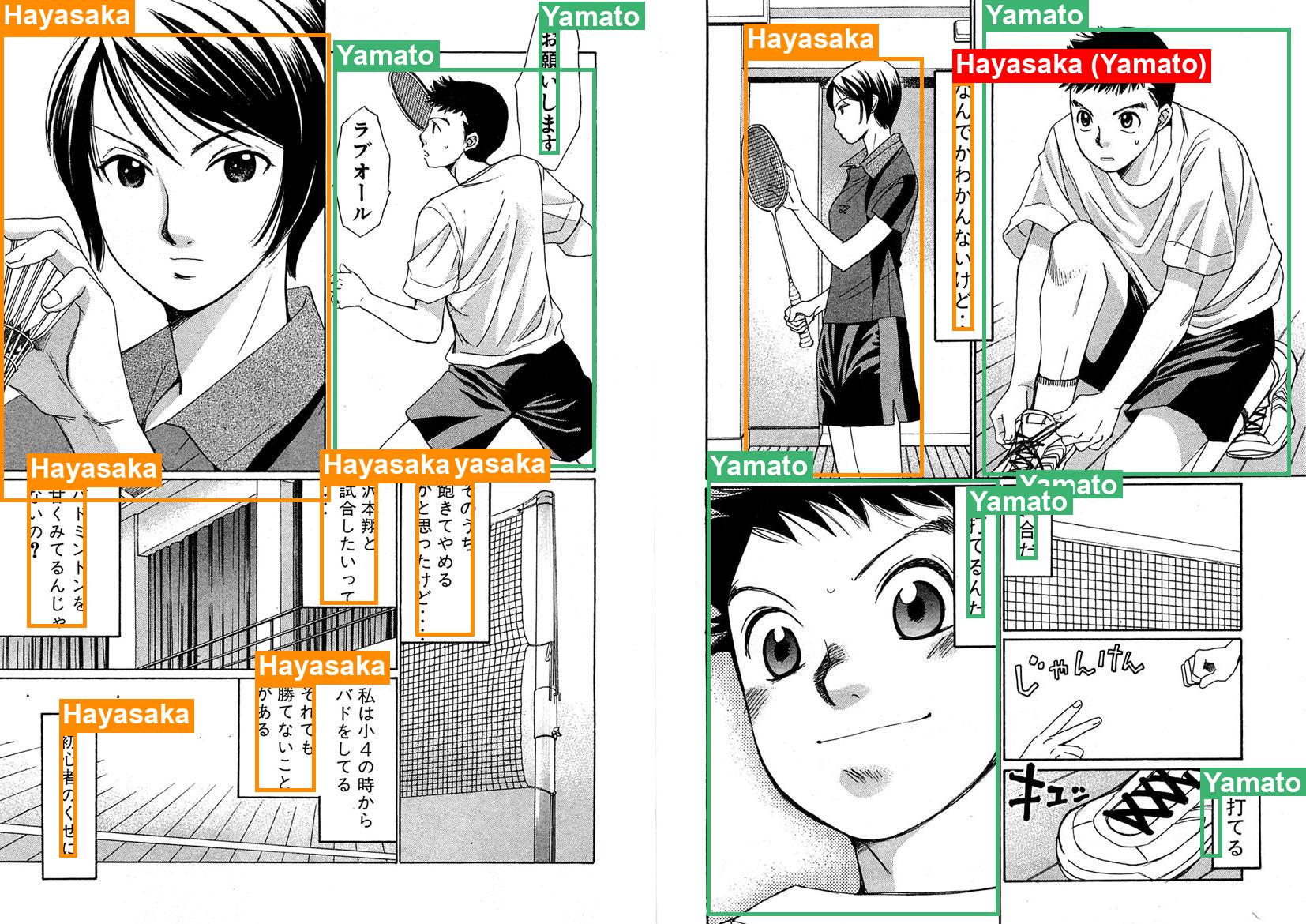}}
    \end{minipage}
    \caption{Results under an entirely zero-shot setting. Courtesy of Saki Kaori.}
    \label{fig:zeroshot}
\end{figure}

Figure~\ref{fig:zeroshot} shows the results under entirely zero-shot settings where only images are provided as inputs.
Experimental settings are described in Section 4.4 of the main paper.

\end{document}